\documentclass[twocolumn,prd,aps,superscriptaddress,preprintnumbers,tightenlines,showpacs,nofootinbib,eqsecnum,amsfonts,amsmath]{revtex4-2}
\usepackage{placeins}
\usepackage{color}
\usepackage{calc}
\usepackage{amsmath,amssymb,graphicx}
\usepackage{tensor}
\usepackage{bm}
\usepackage{times}
\usepackage[varg]{txfonts}
\usepackage[colorlinks, pdfborder={0 0 0}]{hyperref}
\usepackage{float}
\usepackage{dcolumn}
\usepackage[nolist,nohyperlinks]{acronym}
\usepackage{xspace}
\usepackage[abs]{overpic}
\usepackage{pict2e}
\usepackage{enumitem}
\usepackage[usenames,dvipsnames]{xcolor}
\usepackage[utf8]{inputenc}
\usepackage{acronym}
\usepackage{gensymb}
\usepackage{cleveref}
\usepackage{comment}
\usepackage[normalem]{ulem}
\usepackage{longtable}
\usepackage{multirow}
\usepackage{lipsum}
\usepackage{mathtools}
\usepackage{xcolor}

\graphicspath{{figs_main/}}

{\renewcommand{\arraystretch}{1.5}

\usepackage{subfigure}

\usepackage{placeins}

\usepackage{booktabs}

\newcommand{\be}{\begin{equation}}
\newcommand{\ee}{\end{equation}}
\newcommand{\vv}{\mathrm{v}}
\newcommand{\nn}{\nonumber}

\begin{document}
 
 \title{Observing GW190521-like binary black holes and their environment with LISA}

\author{Laura Sberna}
\affiliation{Max Planck Institute for Gravitational Physics (Albert Einstein Institute) Am Mu\"{u}hlenberg 1, 14476 Potsdam, Germany}

\author{Stanislav Babak}
\affiliation{APC, AstroParticule et Cosmologie, 
Université de Paris, CNRS, F-75013 Paris, France}

\author{Sylvain Marsat}
\affiliation{Laboratoire des 2 Infinis - Toulouse (L2IT-IN2P3), Universit\'e de Toulouse, CNRS, UPS, F-31062 Toulouse Cedex 9, France}

\author{Andrea Caputo}
\affiliation{School of Physics and Astronomy, Tel-Aviv University, Tel-Aviv 69978, Israel}
\affiliation{Department of Particle Physics and Astrophysics, Weizmann Institute of Science, Rehovot 7610001, Israel}

\author{Giulia Cusin}
\affiliation{Universit\'e de Gen\'eve, D\'epartement de Physique Th\'eorique and Centre for Astroparticle Physics, 24 quai Ernest-Ansermet, CH-1211 Gen\'eve 4, Switzerland}
\affiliation{Sorbonne Université, CNRS, UMR 7095, Institut d'Astrophysique de Paris, 75014 Paris, France}

\author{Alexandre Toubiana}
\affiliation{Max Planck Institute for Gravitational Physics (Albert Einstein Institute) Am Mu\"{u}hlenberg 1, 14476 Potsdam, Germany}

\author{Enrico Barausse}
\affiliation{SISSA, Via Bonomea 265, 34136 Trieste, Italy \& INFN,
Sezione di Trieste}
\affiliation{IFPU - Institute for Fundamental Physics of the Universe,
Via Beirut 2, 34014 Trieste, Italy}

\author{Chiara Caprini}
\affiliation{Universit\'e de Gen\'eve, D\'epartement de Physique Th\'eorique and Centre for Astroparticle Physics, 24 quai Ernest-Ansermet, CH-1211 Gen\'eve 4, Switzerland} 
\affiliation{CERN, Theoretical Physics Department, 1 Esplanade des Particules, CH-1211 Genève 23, Switzerland}

\author{Tito Dal Canton}
\affiliation{Universit\'e Paris-Saclay, CNRS/IN2P3, IJCLab, 91405 Orsay, France}

 \author{Alberto Sesana}
\affiliation{Department of Physics G. Occhialini, University of Milano - Bicocca, Piazza della Scienza 3, 20126 Milano, Italy}
\affiliation{National Institute of Nuclear Physics INFN, Milano - Bicocca, Piazza della Scienza 3, 20126 Milano, Italy}

\author{Nicola Tamanini}
\affiliation{Laboratoire des 2 Infinis - Toulouse (L2IT-IN2P3), Universit\'e de Toulouse, CNRS, UPS, F-31062 Toulouse Cedex 9, France}

 \begin{abstract}
 Binaries of relatively massive black holes like GW190521 have been proposed to form in dense gas environments, such as the disks of Active Galactic Nuclei (AGNs), and they might be associated with transient electromagnetic counterparts. {The interactions of this putative environment with the binary could leave a significant imprint at the low gravitational wave frequencies
 observable with the Laser Interferometer Space Antenna (LISA).}
 We show that LISA will  be able to detect up to ten 
 GW190521-like black hole binaries, with sky position errors 
 $\lesssim1$ deg$^2$. Moreover, it will measure directly various effects due to 
 the orbital motion 
  around the  supermassive black hole at the center of the AGN, especially the Doppler modulation and the Shapiro time delay.
  Thanks to a careful treatment of their frequency domain signal, we were able to perform the full parameter estimation of Doppler and Shapiro-modulated binaries as seen by LISA.
  We find that the Doppler and Shapiro effects will allow for measuring the AGN parameters ({radius and inclination of the orbit around the AGN, central black hole mass}) with up to percent-level precision.
  Properly modeling these low-frequency environmental effects is crucial to determine  the binary formation history, as well as to avoid biases in the reconstruction of the source parameters and in tests of general relativity with gravitational waves.
  \end{abstract}
    
 \maketitle

\section{Introduction}

In the latest observing run, LIGO and Virgo detected a stellar-mass black hole (BH) binary, GW190521, whose features suggest the binary might have undergone stronger than expected interactions with its environment. 
Firstly,
GW190521 had component masses\footnote{The median values and 90\% credible intervals depend on the assumed priors~\cite{Fishbach:2020qag,Bustillo:2021tga}.} 
$85^{+21}_{-14}\,M_\odot$ and $66^{+17}_{-18}\, M_\odot$~\cite{Abbott:2020tfl, Abbott:2020mjq}, 
with the larger lying squarely within the pair-instability gap $\sim[50,130] M_\odot$ \cite{2002RvMP...74.1015W,2003ApJ...591..288H,2019ApJ...887...53F}.
This has prompted a lively debate in the literature,
with suggestions that the progenitor BHs may have formed not via 
standard stellar evolution in the field, but rather by repeated coalescences in dense environments. The latter include
 globular or nuclear stellar clusters \cite{2017PhRvD..95l4046G,2019PhRvD.100d3027R,2020arXiv200905065F}
or  active galactic nuclei~(AGNs)~\cite{2007MNRAS.374..515L,2020MNRAS.494.1203M,2020ApJ...898...25T}, where BHs may
migrate to and accumulate in the nuclear region at faster pace than
stars due to their larger mass. This in turn enhances their growth by mergers and accretion. These ``dynamical'' formation channels would also explain the apparently large and misaligned spins of the component BHs of   
GW190521~\cite{Abbott:2020tfl, Abbott:2020mjq}.

Secondly, the Zwicky Transient Facility (ZTF) detected
an optical flare (dubbed ZTF19abanrhr) about 34 days
after GW190521, in AGN J124942.3+344929 at redshift $\bar z_s = 0.438$. The position and distance of this system are compatible with
the inferred position and distance of GW190521, especially under the assumption of a uniform prior in mass-ratio~\cite{Bustillo:2021tga}.
Reference~\cite{PhysRevLett.124.251102} then interpreted
ZTF19abanrhr as due to the BH remnant from GW190521
moving in the AGN disk (as a result of the recoil produced
by the anisotropic gravitational wave emission during the merger).
In this picture, the 34-day delay between GW190521 and ZTF19abanrhr
would be due to the time required by the radiation produced at the shock front induced by the recoiling BH remnant to emerge from the surface of the disk. The recoil has been estimated at a velocity $\sim 200$ km/s at $\sim 60$ deg from the midplane
of the disk, whose aspect ratio -- height $H$ to galactocentric radius $a_{\bullet}$ -- was inferred to be $H/a_{\bullet}\sim 0.01$.
Reference~\cite{PhysRevLett.124.251102} also argued that 
the GW190521 binary was likely located 
in a disk migration trap (where
 gas torques cancel out and binaries  accumulate in their inward migration \cite{2016ApJ...819L..17B}), i.e., its distance from 
the nucleus should be $a_{\bullet}\sim 700 GM_\bullet/c^2$, with $M_\bullet\sim10^8 -10^9 \ M_{\odot}$ the mass of the  BH at the center of the AGN.

If indeed GW190521 lived in an AGN disk, it could be the harbinger of a significant population of binary BHs dwelling in gas-rich environments that will be detected in the coming years by ground and space detectors. The large gas densities and the presence of supermassive BHs (SMBHs) at the center of AGN disks could produce observable effects in gravitational wave (GW) signals. For instance, while the gravitational pull of an AGN disk is negligible~\cite{2007PhRvD..75f4026B}, the gravitational interaction of a BH with the sound waves (wakes) that it excites in the surrounding gas leads to effects (dynamical friction and planetary-like migration) that can affect the long-term evolution of a BH binary system~\cite{Barausse:2007dy,Kocsis:2011dr,Barausse:2014tra,Caputo:2020irr}. The surrounding gas can also accrete onto the component BHs, transferring linear momentum to them~\cite{Barausse:2007dy,Barausse:2014tra,Caputo:2020irr}. Moreover, the presence of the central SMBH can significantly affect the waveform, as a result of the accelerated motion of the stellar-origin BH binary around it~\cite{Bonvin:2016qxr,Inayoshi:2017hgw,Tamanini:2019usx}, or because of lensing~\cite{DOrazio:2019fbq,2018MNRAS.474.2975D} and Shapiro time delay~\cite{1964PhRvL..13..789S}. 

{
The most significant among these environmental effects -- accretion, dynamical friction, acceleration -- are believed to be especially prominent at low frequencies, as they appear at negative post-Newtonian (PN) orders in the GW phase relative to the vacuum quadrupole emission~\cite{Barausse:2014tra,Toubiana:2020drf}. While current ground based detectors experience a degradation of their sensitivity at low frequencies (due to seismic noise), environmental effects could be in principle observable with the Laser Interferometer Space Antenna (LISA), by targeting the early inspiral of stellar-origin BH binaries months/years before they merge in the band of ground detectors~\cite{Inayoshi:2017hgw,Tamanini:2019usx,Randall:2019sab,Hoang:2019kye,Deme:2020ewx,Caputo:2020irr,Yu:2020dlm,Toubiana:2020cqv,Toubiana:2020drf}. This opens the possibility to
use the detector to shed light on the physics of AGN disks and on the properties of SMBHs at their centers~\cite{Yu:2020dlm,Toubiana:2020cqv}, as well as to gather important clues on the formation mechanisms of stellar-mass BH binaries~\cite{Amaro-Seoane:2022rxf}. 

On the downside, matter effects could either be degenerate with tests of general relativity (GR) with low-frequency GW signals, or bias the results of such tests if not properly taken into account. Indeed, some modifications of general relativity also predict low-frequency phase contributions, some already partially constrained by LIGO/Virgo observations~\cite{LIGOScientific:2018dkp,LIGOScientific:2021sio}. LISA has the potential to improve low-frequency constraints by about 8 orders of magnitude (in the flux)~\cite{Barausse:2016eii,Toubiana:2020vtf}, provided that matter effects are absent or accurately modelled.
}

In this work, we extend our previous analysis of GW190521-like binaries~\cite{Toubiana:2020drf}, focusing on the modifications to the vacuum waveform induced by the \emph{third body} (the SMBH) and assessing the detectability of its effect with LISA.
{ While motivated by GW190521 and the AGN formation scenario, our results apply more broadly to stellar-mass binaries orbiting a third body with a period comparable to the GW observation time, \emph{in the absence of gas}.}

{Three body systems have a rich and varied phenomenology, depending on the hierarchy of masses and distances of the three bodies. Here, we focus on hierarchical triples in which a compact, circular stellar-mass binary is on a circular orbit around a SMBH with period comparable to LISA's observation time. For these systems, Refs.~\cite{Wong:2019hsq,Randall:2019sab,DOrazio:2019fbq,Yu:2020dlm} found that the Doppler modulation could be measured and used to infer the SMBH mass. If the outer orbit is sufficiently aligned with the line of sight, Refs.~\cite{DOrazio:2019fbq,Yu:2021dqx} showed that lensing could also be detected by LISA. If the binary is even closer to the SMBH, its spin can also be measured thanks to the Lens-Thirring effect~\cite{Yu:2020dlm}, and retro-lensing~\cite{Yu:2021dqx} can be significant. Finally, Refs.~\cite{Hoang:2019kye,Deme:2020ewx,Chandramouli:2021kts} explored other, higher order effects (such as the Kozai-Lidov eccentricity oscillations) and found that these could also be detected by LISA. }

{ These studies made a number of simplifying assumptions, in modelling the Doppler-modulated waveform in the frequency domain with the stationary phase approximation (SPA) and/or by relying on the (sometimes approximate) Fisher matrix to estimate LISA's constraining power, and neglecting the Shapiro time delay. In this work, we address these limitations, and perform realistic parameter estimation for the first time for a binary orbiting a SMBH.
We provide a recipe to obtain the frequency domain waveform for signals strongly modulated by the peculiar motion and the Shapiro delay, {improving on the global SPA commonly used in the literature.} With this {improved} and fast model, we perform a full Bayesian analysis and show that LISA will be able to measure the SMBH mass with percent precision in GW190521-like systems.}

This paper is organized as follows: in Section \ref{sec:rates} we briefly summarize how many (vacuum) GW190521-like binaries LISA could detect, and how well it could measure their vacuum parameters. In Section \ref{sec:theory} we describe how the waveform is modified by the two dominant third-body phenomena: the Doppler and Shapiro effects. We also review other third-body effects, finding them negligible for the system configuration suggested by the ZTF counterpart. Finally, in Section \ref{pe_agn} we model the Doppler and Shapiro-modulated signal in frequency domain and present the results of parameter estimation. We also include some results on the detectability of these signals with vacuum templates. We conclude with some prospects in Section~\ref{sec:concl}.

Throughout our work, we use the following notation: $m_1$, $m_2$ are the masses of the inner (GW190521-like) binary, $\bar z_s$ the cosmological redshift of the source, $M_\bullet$ the mass of the SMBH at the centre of the AGN, $\iota_\bullet$ the inclination of the outer orbit with respect to the line of sight, $a_{\bullet}$ the radius of the outer orbit, $f$ the frequency of the GW. For the outer orbit parameters, see also Fig.~\ref{fig:lensing_setup_new}. We work in units in which $G=c=1$.

\section{Observing (vacuum) GW190521-like binaries with LISA}\label{sec:rates}
 
In this Section, we briefly summarize the prospects of detecting and inferring the properties of GW190521-like binaries with LISA, assuming their vacuum GW signal. 
 
\subsection{Rates and detectability in isolation}
We compute the average number of expected events according to
\begin{equation}\label{eq:barN}
\bar{N} =  R \int {\rm d} \theta {\rm d} \bar z_s \frac{{\rm d} t_{c}}{1+ \bar z_s} {\rm d} m_1 {\rm d} m_2 \frac{{\rm d} V_c}{{\rm d} \bar z_s} p(m_1,m_2)  \, \Theta ({\rm SNR} -{\rm SNR}_{\rm thr}),
\end{equation}
where $\theta$ stands for the angles (inclination, position in the sky and polarization),  $t_c$ is the observed coalescence time, $\bar z_s$ is the source redshift, $m_1>m_2$ the source-frame component masses, $V_c$ the comoving volume, $R$ the intrinsic rate (assumed constant in redshift, details below) and $p(m_1,m_2)$ the mass probability density. We take angles, merger time and comoving volume to be uniformly distributed and assume the Planck 2015 $\Lambda$CDM cosmology~\cite{Ade:2015xua}. We explore different detection thresholds for the signal-to-noise ratio SNR$_{\rm thr}$, mission duration times, instrument duty cycles (SNR$_{\rm DC} = {\rm SNR}_{{\rm DC}=100\%} \times \sqrt{\rm DC}$) and instrument noise curves: ``SciRD", with OMS (optical metrology system) noise of $15$ pm~\cite{LISAscience_document} and ``MRD", with OMS noise of $10$ pm \cite{SciRD}\footnote{{We neglect the LISA galactic confusion noise, as our signals only probe the high frequency part of the instrument sensitivity.}}. In computing the SNR, we neglect the motion of the instrument.
We focus on events between ${\bar z}_{\rm min}=0$ and $\bar z_{\rm max}=0.512$ (corresponding to a comoving distance of $2$ Gpc in our cosmology) and merging within 100 years from the beginning of the LISA mission, as environmental effects can be measured better in events with shorter merger times~\cite{Caputo:2020irr}. 

 \renewcommand{\arraystretch}{1.5}
{\centering
  \begin{table}[t] 
   \begin{tabular}{c|c|c|c}
   noise, $T_{\rm obs}$, DC& GW190521-like & GWTC-3 & GWTC-3    \ massive
   \\
   \cline{1-4}
  
  SciRD, $10 \, {\rm yrs}$, $100\%$  & 
 $7^{+24}_{-7}$ & 
 $22^{+44}_{-17}$  & 
 $5^{+45}_{-5}$\\\hline
 
   SciRD, $6 \, {\rm yrs}$, $100\%$  & 
  $4^{+14}_{-4}$  & 
  $10^{+28}_{-8}$ & 
  $2^{+25}_{-2}$ \\\hline
  
  SciRD, $6 \, {\rm yrs}$, $75\%$  & 
  $2^{+10}_{-2}$  & 
  $6^{+22}_{-5}$ & 
  $1^{+16}_{-1}$\\\hline
  
     MRD, $10 \, {\rm yrs}$, $100\%$  & 
    $13^{+41}_{-13}$  &  
    $70^{+101}_{-47}$  &
    $16^{+73}_{-16}$ \\\hline
      
   MRD, $10 \, {\rm yrs}$, $75\%$  & 
  $11^{+38}_{-11}$  & 
  $43^{+74}_{-29}$ & 
  $10^{+66}_{-10}$ \\\hline
  
  MRD, $6 \, {\rm yrs}$, $75\%$ &
  $6^{+20}_{-6}$  & 
  $19^{+46}_{-15}$ & 
  $5^{+40}_{-5}$\\
   \end{tabular}
\caption{ Average number of GW events from (presumed) AGN binaries detectable by LISA, for different detector noise models, mission durations, duty cycles (DC), and population models. We use an SNR threshold of 8.
}\label{ratesTabel}
\end{table}}

We explore two different ways of predicting the intrinsic rate.
In the first case (``GW190521-like'' in Table~\ref{ratesTabel}) we draw realizations of masses from the posterior distribution obtained by the LIGO/Virgo collaboration (LVC) when analyzing GW190521 with the NRSur waveform model~\cite{Abbott:2020mjq}. We draw the rate from a Gamma distribution whose median and 90\% CI approximately match the values $0.08^{+0.19}_{-0.07} \, {\rm Gpc}^{-3} \, {\rm yr}^{-1}$, the latest LVC estimate for GW190521-like mergers assuming no redshift evolution~\cite{LIGOScientific:2021tfm}. For each rate $R$, we produce samples of the number of events from a Poisson distribution with mean $\bar{N}$ {computed according to Eq.~\eqref{eq:barN}}. Finally, we compute the median and symmetric $90 \%$ confidence interval after sampling over both the rate and the universe realization. The results are shown in Table~\ref{ratesTabel}, and are in agreement with Ref.~\cite{Liu:2021yoy}. We conclude that LISA could detect a few events like GW190521, with a large uncertainty on the actual number depending on the poorly constrained rates and on the actual LISA mission configuration.

{In the second case (``GWTC-3'', in Table~\ref{ratesTabel}), we use the rate estimate (with median $28.1 \, {\rm Gpc}^{-3} {\rm yr}^{-1} $) obtained by the LIGO/Virgo/KAGRA collaboration by analyzing the O1, O2, O3a-b events, assuming a uniform merger rate in comoving volume-time and a ``power law+peak" population model~\cite{LIGOScientific:2021psn}.
We sample from the posterior distribution of the power law+peak model parameters provided in Ref.~\cite{LIGOScientific:2021psn}. 
For each population model realization we estimate the mean number of events, and produce realizations from a Poisson distribution with mean $\bar{N}$. Finally, we compute the median and symmetric $90 \%$ confidence interval and report them in Table~\ref{ratesTabel}.}

{
The formation mechanism of the GWTC-3 events is still being debated, with one Bayesian analysis finding for about 25\% of them a preference for an AGN origin (in a specific model)~\cite{Gayathri:2021xwb}. 
To mimic the AGN sub-population, we consider the sub-set of events with primary masses above $45 \, M_{\odot}$ (the column named ``GWTC-3 massive" in Table~\ref{ratesTabel}).

Our predictions for the overall rate of stellar-mass binaries detectable by LISA lie between a few and $\mathcal{O}(100)$, and are broadly consistent with other estimates in the literature once we account for differences in the rate, detection threshold, etc. (see, e.g., ~\cite{Moore:2019pke,Toubiana:2022vpp}).\footnote{Note that our rates exclude stellar-mass binaries on wider orbits (i.e., merging in more than $100$ years) that LISA could detect in our Galaxy, as predicted by population synthesis studies~(see \cite{Wagg:2021cst} and references therein).}
Furthermore, based on the latest catalog, we predict there will be $\mathcal{O}(10)$ events detected by the LISA mission likely to occur \emph{in AGNs} (still with a considerable uncertainty).
}
Note that, within the detectable events in any of the configurations in Table~\ref{ratesTabel}, only a fraction will be realistic (merging within 10 years) multiband events.

\begin{figure*}[!htb]
\centering
\includegraphics[width=.98\columnwidth]{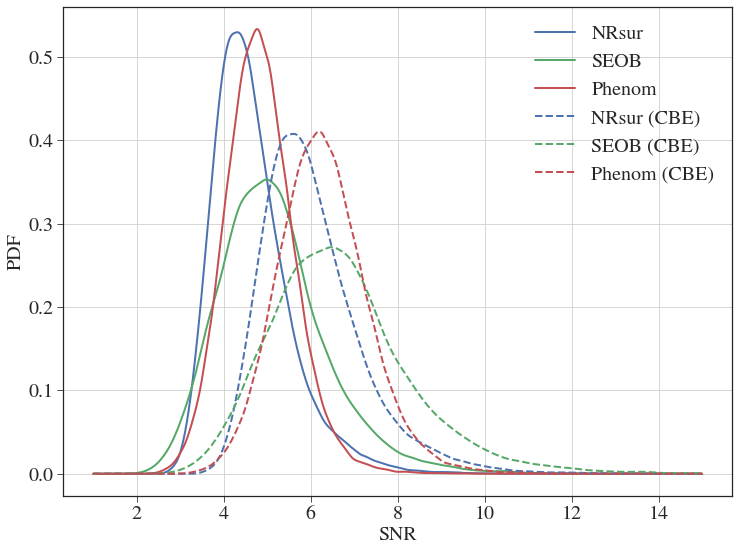}\hspace{.5cm}
\includegraphics[trim={0.pt 3.5pt 0pt 0.pt},width=.96\columnwidth]{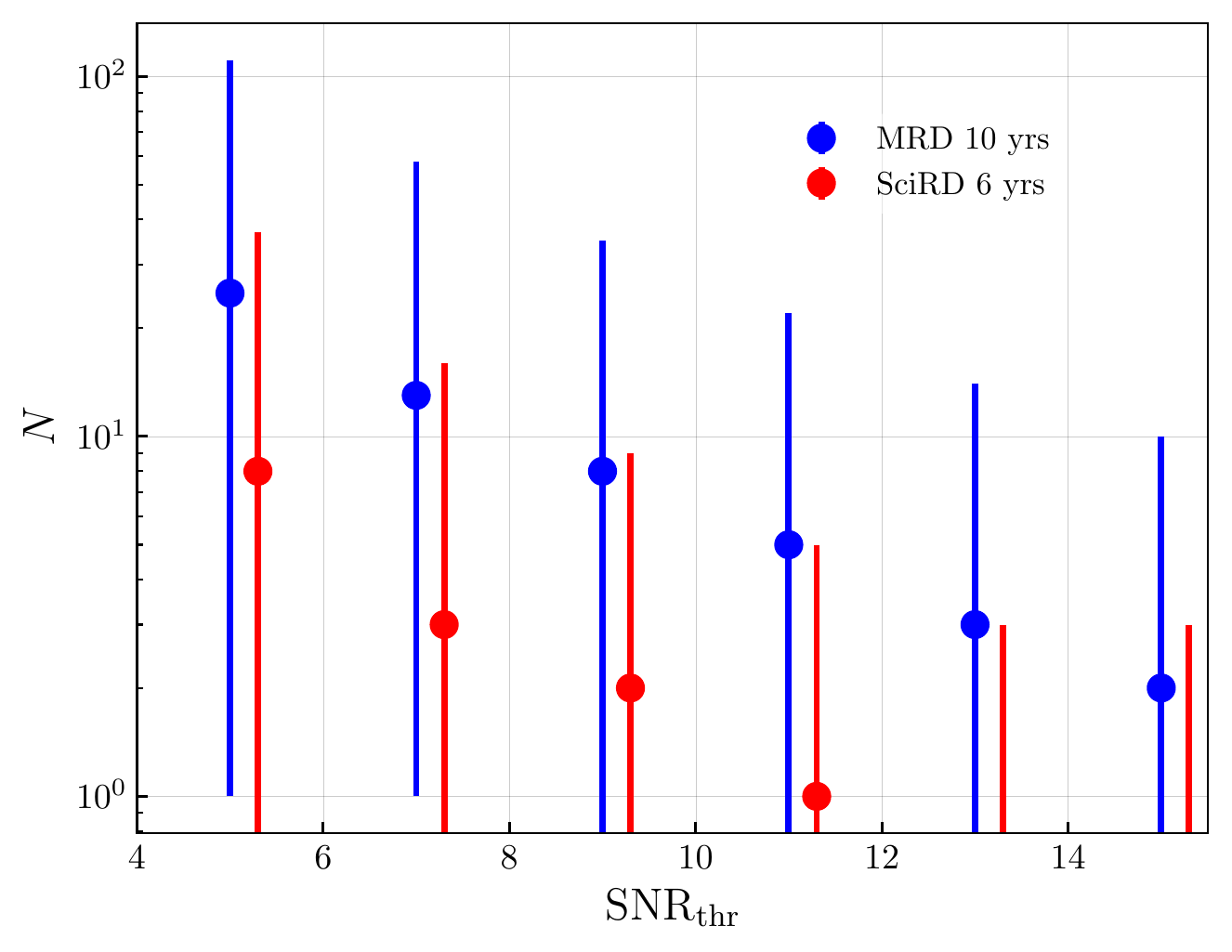}
\caption{
Left: distribution of the SNR in LISA for GW190521, averaged over the sky position and polarization assuming that the binary merges shortly after 10 years of LISA observations. Solid lines correspond to SciRD detector sensitivity, while dashed lines corresponds to the current-best-estimate (CBE) sensitivity, similar to MRD at high frequencies. Right: number of GW190521-like events detectable by LISA as a function of the SNR threshold, for a fiducial (SciRD, 6 yrs) and optimistic (MRD, 10 yrs) mission configuration. SciRD points are shifted for clarity. Both panels assume a conservative duty cycle of 75\%.}\label{SNR_GW190521}
\end{figure*}

In Fig.~\ref{SNR_GW190521} we also show the distribution of the 10-year SNR of GW190521-like binaries in LISA, averaged over the sky position and polarization angle, and with merger time of $10$ yrs. Figure~\ref{SNR_GW190521} also reports the number of GW190521-like events detectable by LISA as a function of the SNR threshold, for a $75\%$ duty-cycle and a fiducial (SciRD, 6 yrs) and optimistic (MRD, 10 yrs) mission configuration. 

Overall, the SNR distribution in Fig.~\ref{SNR_GW190521}, as well as the rates in Table~\ref{ratesTabel}, point to the fact that GW190521-like binaries will be hard to detect with LISA, unless found at a closer distance. 
Detection prospects would improve with strategies (such as multiband detections~\cite{Moore:2019pke}) requiring a lower SNR threshold, as seen in Fig.~\ref{SNR_GW190521}, {right panel.}

 \subsection{Parameter estimation in isolation}
 
Despite the large masses of GW190521-like systems, (as defined by the samples released by the LVC), the morphology of these GW signals is very similar to the one of stellar-mass BH binaries, i.e., we observe the system during its early inspiral and it then leaves the LISA band as it starts chirping. We therefore refer to Ref.~\cite{Toubiana:2020cqv} for a complete description of the correlations between parameters and the accuracy in their measurement.
 Among intrinsic parameters only the chirp mass can be precisely measured, with relative error $10^{-4}$. Thanks to its long duration in the LISA band, the source can be very well localised, within $1 \; {\rm deg}^2$, see also Ref.~\cite{Liu:2021yoy}. Finally, the distance to the source is measured with a $60\%$ error. As we will see in Section \ref{pe_agn}, the measurement of the binary parameters barely changes when it is near a SMBH. 
 \section{GW190521 around an AGN}\label{sec:theory}
\begin{figure}[t]
\includegraphics[trim={3.cm 7.cm 2.5cm 4.cm},clip,width=0.99\columnwidth]{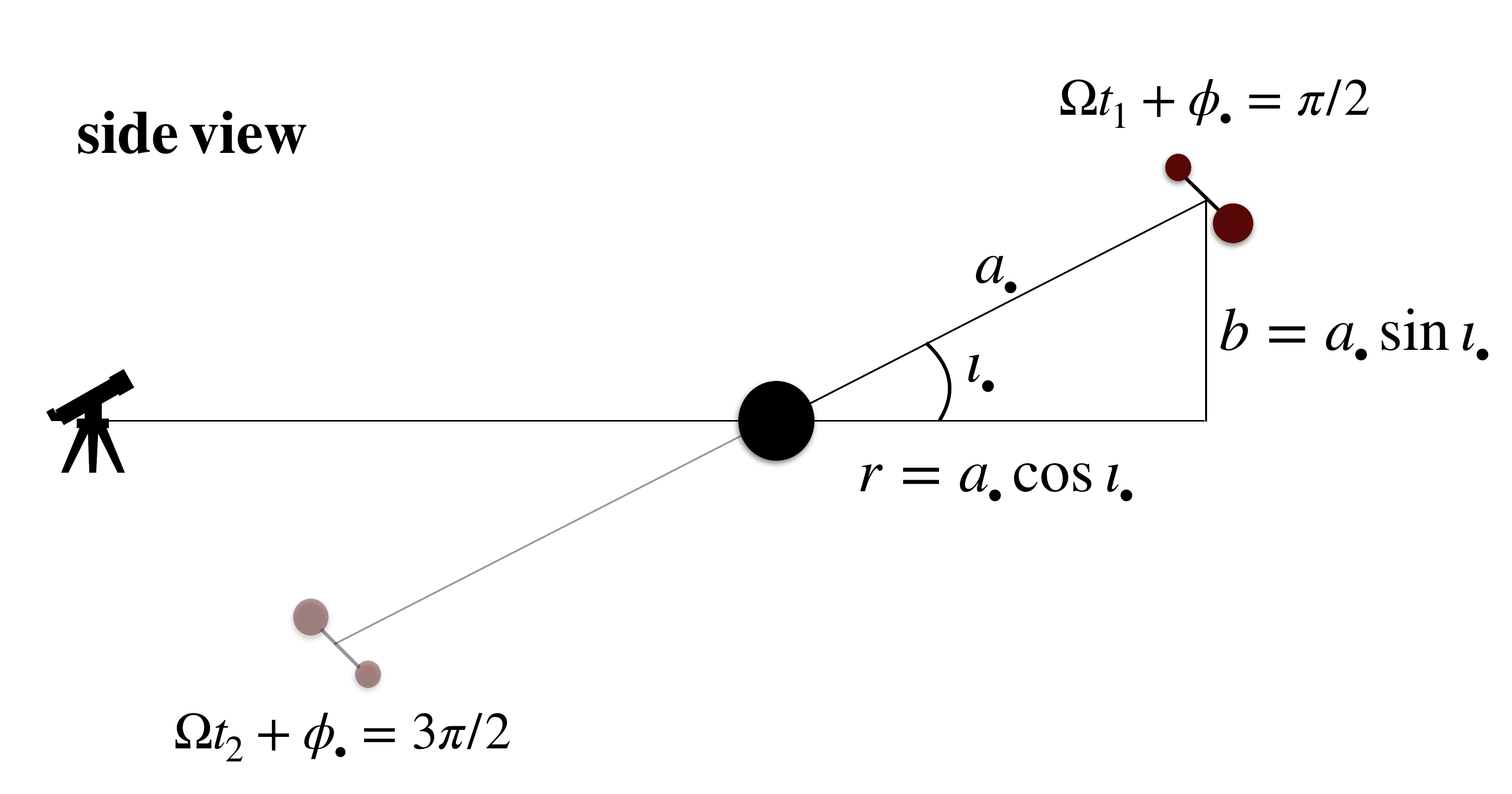}
\caption{Snapshot of the system when the inner binary is behind (at $t=t_1$) or in front (at $t=t_2$) of the SMBH, i.e., when the source, the central black hole and the observer all lie in the page plane. The outer orbit is perpendicular to this plane.} \label{fig:lensing_setup_new}
\end{figure}

In Ref.~\cite{Toubiana:2020drf} some of the current authors assessed the detectability of deviations from purely vacuum waveforms arising from AGN \emph{matter} effects, such as gas accretion and dynamical friction. If GW190521-like events occur in dense gaseous environments, with sizable values for the gas density, $\rho_{\rm gas} \simeq 10^{-10} \, \rm g/cm^3$, and accretion rates, $f_{\rm Edd} \simeq \mathcal{O}(1)$, then LISA will be able to detect these effects in the GW signal. 
In the following, we will focus our attention on the AGN environmental effects related to the presence of a \emph{third body}, which have been not analyzed in depth in previous works. 

We will focus on binaries at 1) intermediate distances from the SMBH, with $a_{\bullet}=\mathcal{O}(100) M_{\bullet}$, and 2) with generic inclinations between the outer orbit and the line of sight. Because of the putative identification with an electromagnetic counterpart~\cite{PhysRevLett.124.251102}, we also refer to these binaries as GW190521-like.
The two dominant effects in these conditions are the \emph{Doppler} and \emph{Shapiro effects}, which we describe in detail in this Section. We will only briefly discuss effects that might be relevant at shorter separations (e.g.~Lense-Thirring precession) or for highly aligned systems (lensing).

Our choice for the distance from the SMBH is motivated by the fact that we expect AGN binaries to be preferentially located in disk traps~\cite{Yang2019,Secunda:2018kar}, where inward and outward disk torques balance and black holes can accumulate and merge hierarchically. These traps are typically found between $\mathcal{O}(10)M_\bullet$ and $\mathcal{O}(10^3)M_\bullet$ from a central SMBH of mass $M_\bullet$~\cite{Bellovary:2015ifg} (although, see also~\cite{Pan:2021ksp}), and could even occur close to the innermost stable circular orbit~\cite{Peng:2021vzr}. 

We also assume that the outer orbit is circular, as migrating bodies in AGN disks, similarly to planets in protoplanetary disks, are expected to circularize~\cite{TanakaWard2004,Cresswell2008}. Finally, at distances $\mathcal{O}(100) M_{\bullet}$ from the central object, we can assume that the SMBH is nonrotating (spin-related effects are negligible, see Section~\ref{sec:other}). 

For the inner binary, we restrict our analysis to a spin-aligned and circularized system, leaving the modelling of spin-induced precession and eccentricity in binaries around a SMBH {(see, e.g.,~\cite{Liu:2020tcd,Samsing2022})} for future studies.

\subsection{The waveform in the observer frame}
We now derive the waveform in the observer frame (``o") and relate it to the usual expression in the source frame (``s"). The relation between the observer-frame and source-frame time, as well as the observed redshift, can be derived as follows. We use the McVittie metric~\cite{McVittie:1933zz,Minazzoli:2019ugi}
\begin{equation}\label{eq:McV}
    ds^2 = -\left[ \frac{1 -\mu(t,r)}{1+\mu(t,r)}\right]^2 dt^2 + \left[ 1+\mu(t,r)\right] a^2(t) d\vec{x}^2 \; ,
\end{equation}
with $\mu (t,r) = m/(2 r a(t))$
to describe an observer in an expanding (flat) Universe in the presence of a static object of mass $m$ (corresponding to a Schwarzschild BH when $a(t)=$1), from which the coordinate $r$ originates. This is a good description for sources at cosmological distances ($\gtrsim$ Gpc) from us -- for those at closer distances, we could simply use the Schwarzschild metric.
For $m / r \ll 1$, Eq.~\eqref{eq:McV} can be put in the standard form of a perturbed expanding universe in longitudinal gauge (see, e.g.,~\cite{Durrer:2020fza}),
\begin{align}
    ds^2 =& -(1+2 \Psi)dt^2 + a^2(t)(1-2\Phi)d\vec{x}^2 \nonumber \\
    =& \,a^2(\tau)\left[
    -(1+2 \Psi)d\tau^2 + (1-2\Phi)d\vec{x}^2\right] \, ,
\end{align}
with $\Psi = \Phi = - 2 \mu$. 

The source's four-velocity is $u^{\mu}_s =\frac{\gamma_s}{a}(1-\Psi_s, \vec{v}_s)$ with $\gamma_s = (1-|\vec{v}_s|^2 + \Psi_s^2+ \mathcal{O}(\Psi_s^3,|\vec{v}_s|^2 \Phi_s))^{-1/2}$ and we neglect the motion and the potential at the observer, which 
can be taken into account separately, in the GW detector response. 
At the source, the wave-vector of the gravitational wave,  perturbed by the presence of the point mass (i.e.~the central BH), is $k_s^{\mu} =a^{-2} (1+ \delta k_s^0 , \hat{n}+ \delta \vec{k}_s)$ with $\hat{n}$ pointing from the source to the observer, and where we again neglect  perturbations at the observer. Because we assumed a circular outer orbit, the magnitude of the source velocity is constant in time, $\gamma_s={\rm const}$.
The redshift between the observer and the source is then given by~\cite{Bonvin:2005ps,Durrer:2020fza}
\begin{align}\label{eq:redshift_all}
(1+z) = \frac{dt_o}{dt_s} = \frac{(g_{\mu\nu}u^{\mu}k^{\nu})_s}{(g_{\mu\nu}u^{\mu}k^{\nu})_o}\simeq& \gamma_s \frac{a_o}{a_s} \left( 1 - \vec{v}_s \cdot \hat{n} + \Psi 
\right)  \\ \nonumber
= & (1+\bar{z}_s)(1 + z_D(t_s) + z_S(t_s)) \; ,
\end{align}
where we neglected higher order terms mixing the potential and the peculiar motion, as well as the time derivatives of the potentials. {We also absorbed the constant $\gamma_s$ factor in the cosmological redshift $\gamma_s \, a_o/a_s \rightarrow (1+\bar{z}_s)$.} From Eq.~\eqref{eq:redshift_all}, we see that the observed redshift receives contributions from the cosmological redshift $\bar{z}_s$, the peculiar motion (Doppler effect, $z_D$) and the gravitational influence of the third body (Shapiro effect, $z_S$). 

We can integrate the redshift to obtain the time measured by the observer. We neglect the change in the expansion of the universe during the time the signal is propagating from the source to the observer, so that $1+\bar z_s={\rm const}$ (for works taking into account the variation of the background expansion, see \cite{Seto:2001qf,Nishizawa:2011eq,Bonvin:2016qxr,Tamanini:2019usx}).
Integrating from a reference time that we set to zero for simplicity, we find
\begin{align} \label{eq:timing}
    t_o = & (1+\bar z_s) \left[ t_s -\int_0^{t_s} \vec{v}_s \cdot \hat{n}\, dt'
    + \int_0^{t_s} \Psi_s
     \,dt' \right] \nonumber \\
    =  & (1+\bar z_s) \,t_s + d_D(t_s) + d_S(t_s)\, \nonumber \\
    \simeq & t_{s,\bar z} + d_D((1+\bar z_s)^{-1}t_{s,\bar z}) + d_S((1+\bar z_s)^{-1}t_{s,\bar z}) \, ,
\end{align}
{where we define $dt_{s,\bar z} \equiv  (1+\bar z_s) \,dt_s $.} In the last line of Eq.~\eqref{eq:timing}, we expressed the delays in terms of $t_{s,\bar z}$, which takes into account the background cosmology. 
This is a convenient step for computing the waveform in the observer frame, as we will use waveform models that already take into account {the standard} propagation in a cosmological background.
The second term in Eq.~\eqref{eq:timing} represents the delay due to the peculiar motion (Doppler delay, in short, Section~\ref{S:orb_effects}), while the third term is the gravitational delay due to the central mass (Shapiro delay, Section~\ref{sec:Shapiro}).

Finally, the GW signal emitted by the binary in orbit around the SMBH and seen by our detector can be written as 
$h_o(t_o) = A_o(t_o) \, e^{i \varphi_o(t_o)} $. {Neglecting amplitude corrections, the observed signal is related to the signal in the source frame by a transformation of the argument,} 
\begin{align} 
A_o(t_o)=& 
A_s(t_{s,\bar z})  + \mathcal{O}(|\vec{v}_s| A_s, \Psi_s A_s) \simeq A_s(t_{s,\bar z}) \, , \label{eq:gw_ampl} \\
\varphi_o(t_o)=& \varphi_s(t_{s,\bar z}) \, .\label{eq:gw_phase} 
\end{align}
{Here $A_s$ and $\varphi_s$ are the amplitude and phase of the GW signal taking into account the propagation from the source to the observer in the cosmological background (i.e., they are expressed in terms of the luminosity distance and redshifted masses). 
The observed time $t_o$ is given by Eq.~\eqref{eq:timing}. 

In this work, we neglect the amplitude corrections the GW receives in the observer frame from the peculiar motion {and the potential at the source, which are of the order} $(1+z_{D,S}(t_s))$}~\cite{Boyle:2015nqa,DOrazio:2019fbq}. In other words, we only account for the transformation in the argument of the waveform (i.e., the time shift). We justify neglecting amplitude corrections in the sections to follow.
We also focus on the $l=m=2$ mode, and neglect the mode mixing generated by peculiar motion~\cite{Boyle:2015nqa,Torres-Orjuela:2018ejx,Torres-Orjuela:2020dhw}. 

In the remainder, we describe in more detail the two effects contributing to the time shift~\eqref{eq:timing}. For simplicity, we drop the subscript in the source-frame time, setting $t_{s,\bar z}=t$.

 \subsection{The Doppler effect}
 \label{S:orb_effects}
 
 \begin{figure*}[t]
\centering
\includegraphics[trim={0.3cm 0.cm 0cm 0cm},clip,width=.95\columnwidth]{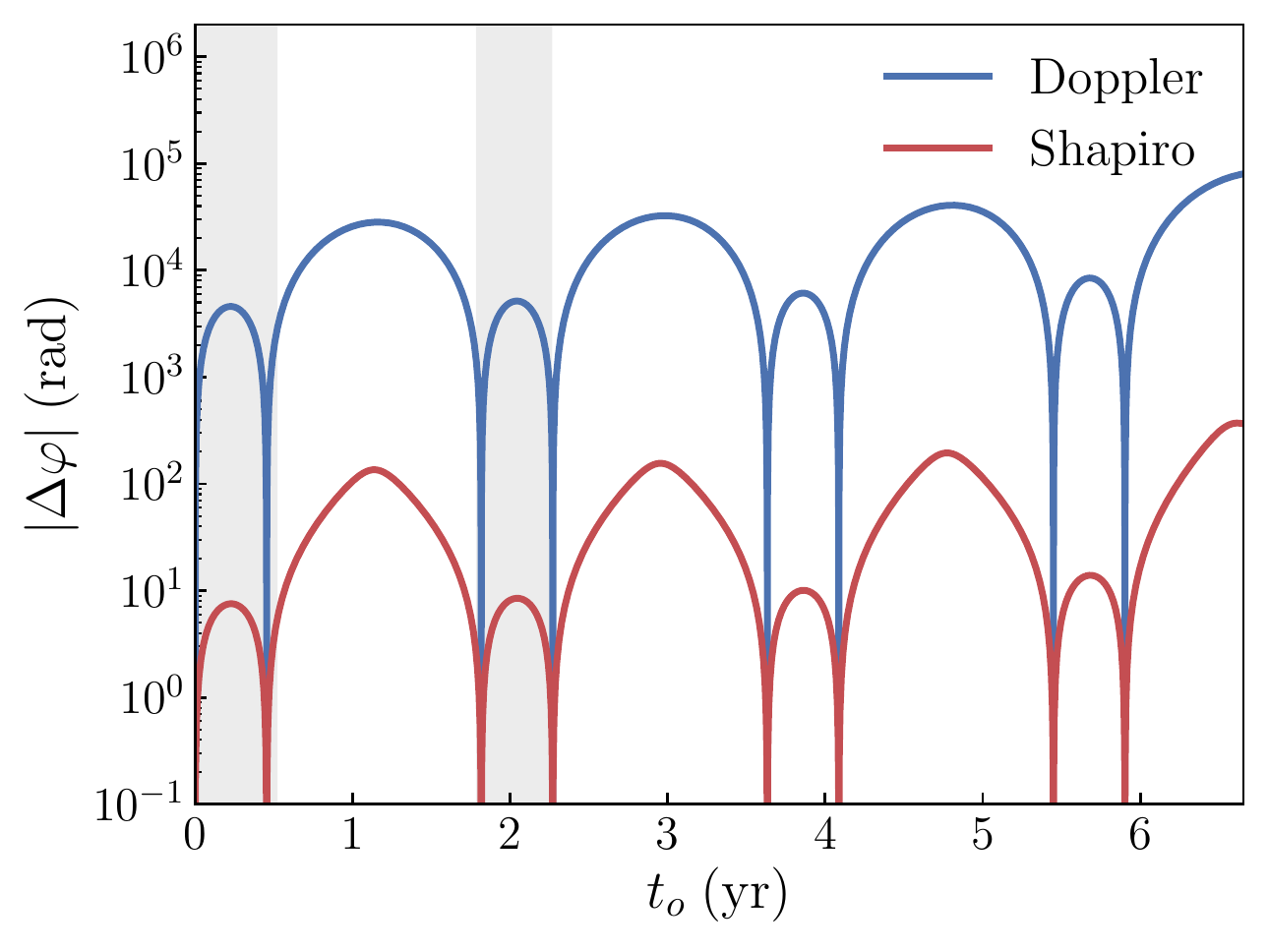}
\hspace{0.5cm}
\includegraphics[trim={0.3cm 0.cm 0cm 0cm},clip,width=.95\columnwidth]{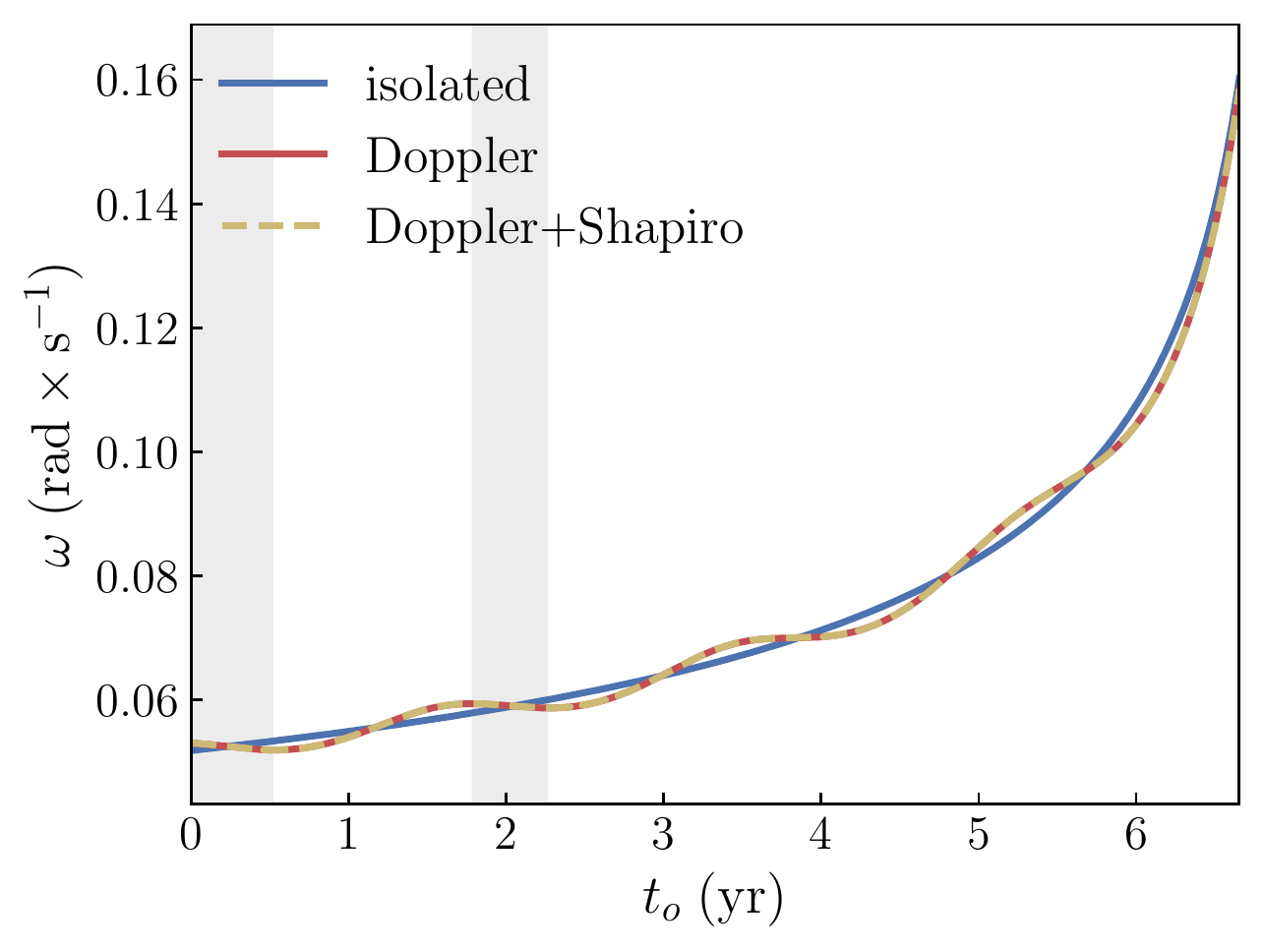}
\caption{{\it Left:} The dephasing between the signal emitted by a GW190521-like binary in isolation and one affected by the Doppler (blue) or Shapiro (red) effect. 
{\it Right:} the derivative of the GW phase $\omega = \frac{d}{dt}\varphi(t+d_{\rm effect}(t))$ in isolation and in the presence of the Doppler and Shapiro effects. 
In both figures, the shaded bands indicate the antichirping phases of the Doppler-shifted signal, $\frac{d}{dt}\omega <0$ and the outer orbit has a \emph{large} inclination with respect to the line of sight, $\iota_\bullet = \pi/6 \, {\rm rad} = 30 \, {\rm deg}$. The other outer binary parameters are set to $M_{\bullet,\bar z} = 10^8 M_{\odot}$, $a_{\bullet, \bar z} = 700 M_{\bullet, \bar z}$, $\phi_{\bullet} = -3\pi/4$. The inner binary parameters are set to the median values of GW190521, with merger time $t_{c,\bar z}= 7 \; {\rm yr}$. }
\label{fig:plot_DopShaonly}
\end{figure*}

The leading order effect of the outer orbit is a variation of the distance between the detector and the source, causing a varying Doppler delay $d_D(t)$.

The signal in the observer frame is given by Eqs.~\eqref{eq:gw_ampl}, \eqref{eq:gw_phase}, with $t_o = t + d_D(t)$. 
The delay
$d_D(t)$ is the time dependent change in the arrival time of the signal compared to a source with negligible proper motion:
from the second term in Eq.~\eqref{eq:timing}, one gets ({recalling that $t$ stands now for $t_{s,\bar z}$})
\begin{equation}\label{eq:DDT}
    d_D(t) = -(1+\bar z_s)\int_0^{ \frac{t}{1+{\bar{z}_s}}} \vec{v}_s \cdot \hat{n}\, dt' = (1+\bar z_s) r(t) + {\rm const.},
\end{equation}
where the distance $r(t)$ is the projection onto the line of sight (pointing now from the observer to the source) of the outer orbit, which can be written as 
\be
r(t) = a_{\bullet} \cos \iota_\bullet \sin ( \Omega_{\bar z} t +\phi_{\bullet})\, ,
\label{eq:rt}
\ee
with the system geometry defined in Fig.~\ref{fig:lensing_setup_new} and $\Omega_{\bar z}=M_{\bullet, \bar z}^{1/2}a_{\bullet, \bar z}^{-3/2}$. As expected, the time delay $d_D$ only depends on the redshifted environmental parameters: $a_{\bullet, \bar z}$, $M_{\bullet, \bar z}$. 
The constant in Eq.~\eqref{eq:DDT} can be re-absorbed in the definition of the time of coalescence.

From Eq.~\eqref{eq:DDT}, we see that the Doppler delay has a mild dependence on the inclination of the outer orbit, $\iota_\bullet$. 
We also see that the distance from the SMBH enters the waveform in combination with the orbital inclination, in the factor $a_{\bullet} \cos \iota_{\bullet}$. The SMBH mass also only enters in combination with the orbital separation, in the orbital frequency $\Omega$. 
When the Doppler effect is the only orbital effect measured in the GW signal, these two combinations {of the orbital parameters} are the only observable quantities.

At a typical disk trap location, the observed outer orbital period is
\be \label{eq:outer_period}
T=2 \,\text{yr} \left(\frac{a_{\bullet, \bar z}}{700 \, M_\bullet}\right)^{3/2} \left(\frac{M_{\bullet,\bar z}}{10^8 M_{\odot}}\right) \,,
\ee
which is comparable with the LISA mission duration, and we expect the delay to induce appreciable variations in the GW phase.
{An estimate of the amplitude of the phase modulation induced by the Doppler effect is given by}
\begin{align}\label{eq:mod_Doppler}
    \delta \phi_D &\simeq 2\pi f d_D \simeq 2\pi f a_{\bullet, \bar z} \cos \iota_\bullet \nonumber \\
    &\simeq 2 \times  10^4 \; \mathrm{rad} \; \cos \iota_\bullet \left(\frac{f}{10\mathrm{mHz}}\right) \left(\frac{a_{\bullet, \bar z}}{700M_{\bullet,\bar z}}\right) \left(\frac{M_{\bullet, \bar z}}{10^8 M_\odot}\right) \, .
\end{align}
This phase modulation is, at realistic orbital separations, the largest non-vacuum effect in AGN binaries, much larger than the phase modulation of the LISA spacecraft motion itself (approximately~$30$ rad). The large dephasing induced by the peculiar motion in time domain can be seen in Fig.~\ref{fig:plot_DopShaonly}.\footnote{ Figure~\ref{fig:plot_DopShaonly} shows that the dephasing induced by the Doppler (and Shapiro) effect is not symmetric over an outer orbit. Indeed, the dephasing can be approximated as $\Delta \phi = \phi(t+d(t)) - \phi(t) \simeq \dot\phi(t)d(t)$, where $d(t)$ is symmetric around an orbit, and $\dot\phi(t)$ is not (it grows monotonically with time).} The detectability of the peculiar motion phase shift was explored, for more moderate accelerations, in several studies~\cite{Bonvin:2016qxr,Randall:2019sab,Tamanini:2019usx,Wong:2019hsq}.

{As anticipated in Ref.~\cite{Toubiana:2020drf}, for orbital periods comparable or smaller than the LISA observational time, the Doppler effect modulates the signal to the point that the GW frequency evolution is not monotonic during observations, see Figs.~\ref{fig:plot_DopShaonly} and \ref{fig:plot_DopSha}. Because the time-to-frequency map is multivalued, a global use of the SPA (appearing e.g.~in \cite{Yu:2020dlm,Yu:2021dqx}) would not be adequate to describe the signal in the frequency domain. In this work, we will focus on this regime and explain how to obtain a more accurate frequency-domain waveform (Section~\ref{pe_agn}). }

The Doppler effect will also modulate the GW amplitude through a time dependent redshift factor. From Eq.~\eqref{eq:redshift_all}, {and keeping track of the redshift factors in the transformation from the source-frame time,} 
\begin{align}
z_D(t) =& (-\vec{v}_s \cdot \hat{n}) = (1+\bar{z}_s) \dot{r}(t)  \\ \nonumber
=& a_{\bullet, \bar z} \Omega_{\bar z} \cos \iota_\bullet \cos ( \Omega_{\bar z} t +\phi_{\bullet})\, .
\end{align} 
This will induce a modulation of the GW amplitude similar to the modulation of the GW frequency, {with amplitude $\delta z_D = z_{D,\rm max} - z_{D,\rm min}$} of order
\be
\delta z_D \simeq 0.04 \, \cos \iota_\bullet \, \left(\frac{a_{\bullet, \bar z}}{700M_{\bullet, \bar z}}\right)^{-1/2} \left(\frac{M_{\bullet, \bar z}}{10^8 M_\odot}\right)^{1/2} \, .
\ee
Gravitational wave detectors are more sensitive to phase, rather than amplitude, modulations. For reference, in the case of a GW190521-like event at $1.5$ Gpc distance, LISA will measure the luminosity distance with precision of order $50 \%$, see Section~\ref{pe_agn}, Fig.~\ref{F:ZTF_gr_pe}.   A Doppler-induced amplitude modulation of percent level will thus be negligible and in any case subdominant compared to the phase modulation induced by the same effect, {as we confirm numerically in Section~\ref{pe_agn}.} {We therefore neglect the amplitude modulation when performing parameter estimation.} 

For sufficiently large outer orbital separations, i.e., for orbital periods much longer than the LISA observational time, the peculiar motion reduces to a constant peculiar velocity and a constant centripetal acceleration, see Appendix~\ref{app:peculiar}. 
In this case, the constant peculiar velocity can be reabsorbed in the constant redshift (and the latter in the redshifted chirp mass) and the centripetal acceleration produces a single {-4PN} term in the GW phase.
Parameter estimation in this limit was discussed in Ref.~\cite{Toubiana:2020drf} (see also \cite{Tamanini:2019usx}).

\subsection{The Shapiro effect}\label{sec:Shapiro}

\begin{figure*}[!htb]
\centering
\includegraphics[trim={0.1cm 0.cm 0.2cm 0cm},clip,width=.95\columnwidth]{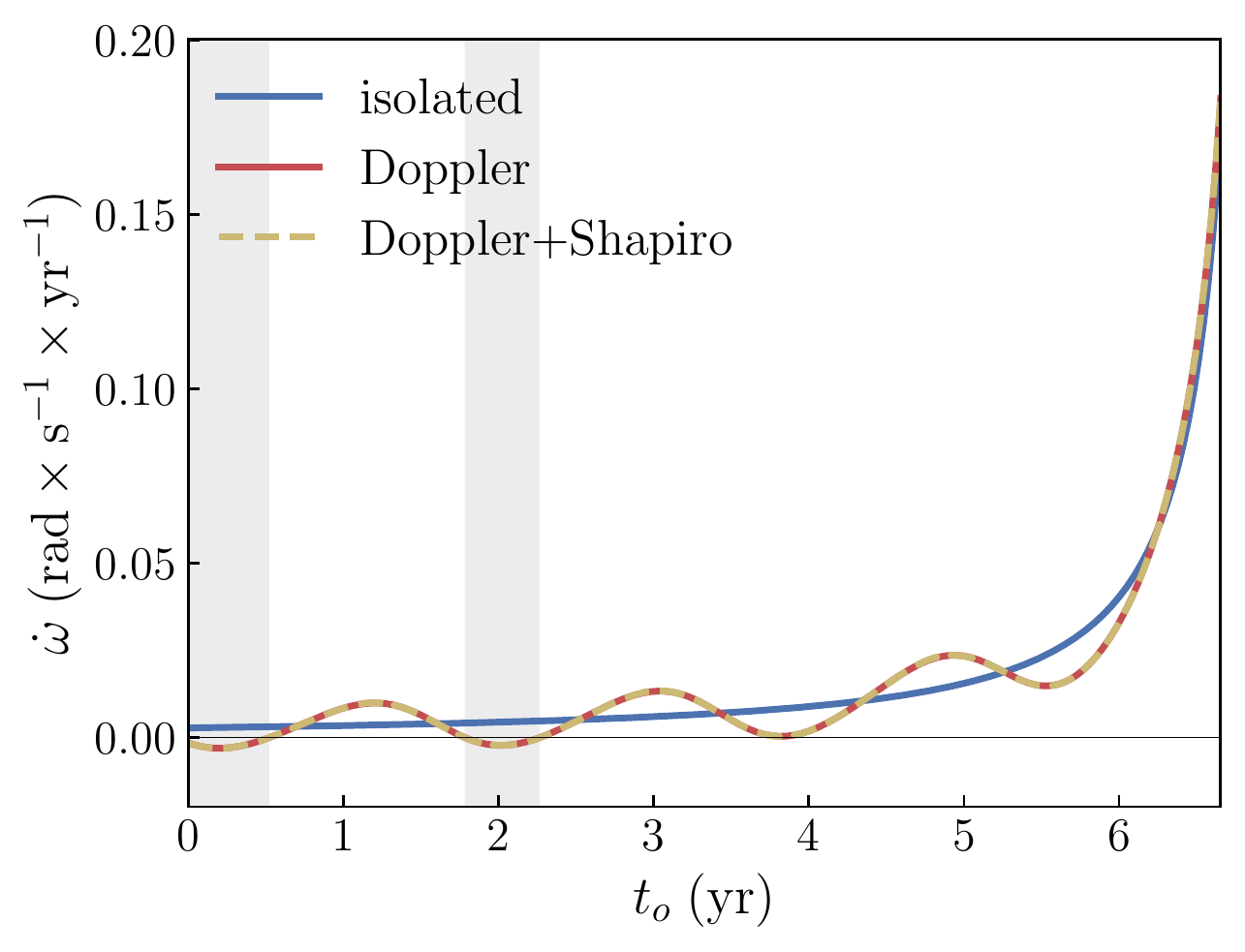} 
\hspace{0.5cm}
\includegraphics[trim={0.1cm 0.cm 0.2cm 0.2cm},clip,width=.95\columnwidth]{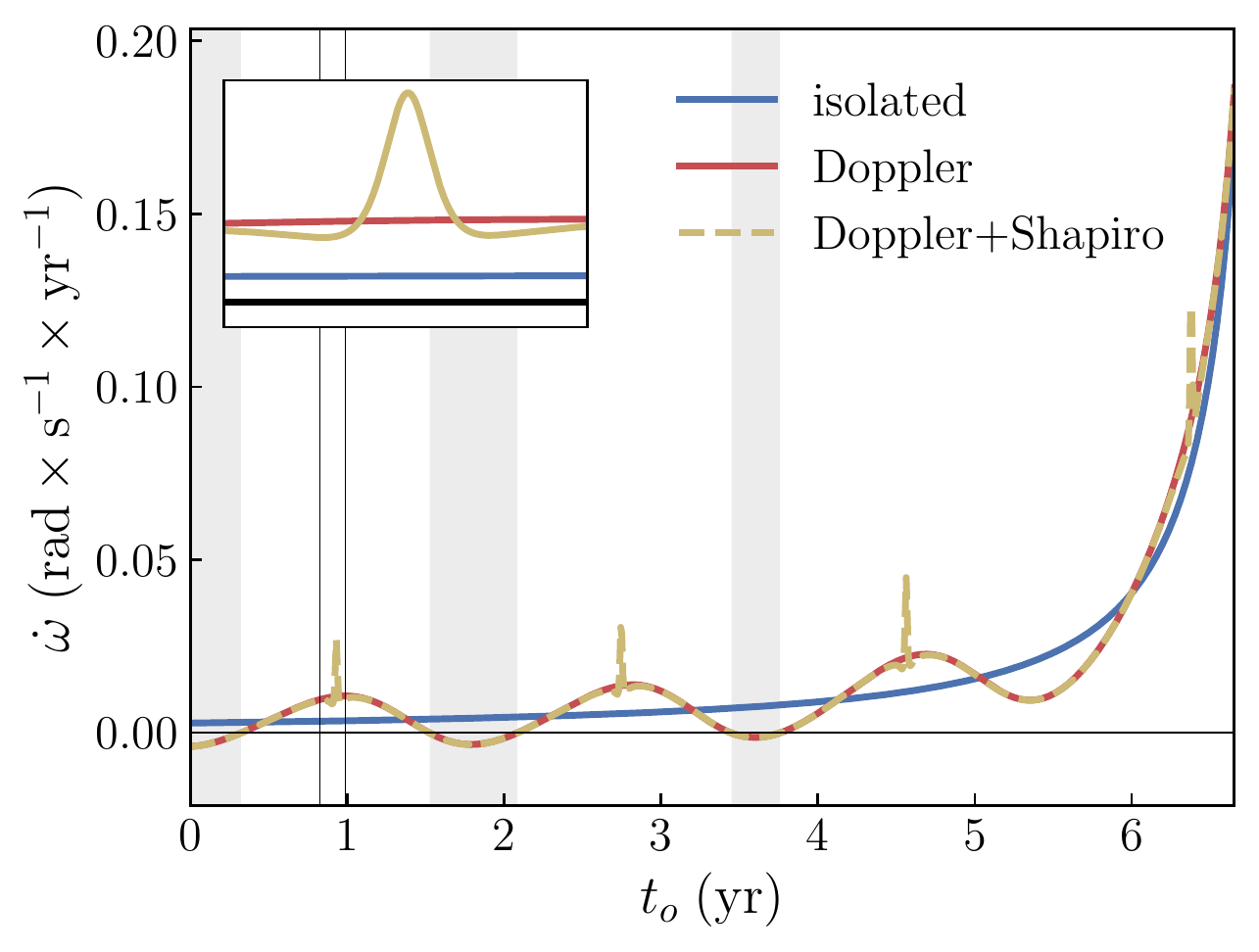}
\caption{ The second derivative of the GW phase $\dot{\omega} \equiv \frac{d}{dt}\omega = \frac{d^2}{dt^2}\varphi(t+d_{\rm effect}(t))$ in isolation and in the presence of the Doppler and Shapiro effects. 
We compare a system with large ($\iota_\bullet = \pi/6 \, {\rm rad} = 30 \, {\rm deg}$, left) and small ($\iota_\bullet = -0.05 \; {\rm rad} \simeq - 3 \; {\rm deg} $, right) outer orbit inclination. The remaining system parameters and plot features are the same as in Fig.~\ref{fig:plot_DopShaonly}. 
The figure on the right includes a zoom-in of the region marked by vertical lines, corresponding to a month around the moment the source is aligned behind the SMBH. At each passage, the Shapiro effect is strongly enhanced in the small inclination case. Note that at small inclinations the effect of lensing, not included here, would also be significant.}
\label{fig:plot_DopSha}
\end{figure*}

The second most relevant effect in our setting is the Shapiro effect, due to the non-vanishing gravitational potential of the third body.

The Shapiro time delay is the delay in the time of arrival of a signal (with respect to the signal traveling in {flat spacetime}) due to the non vanishing gravitational potential along the line of sight. Assuming the central object is described by the Schwarzschild metric\footnote{Note that, while Eq.~\eqref{eq:shapiro} is derived assuming the Schwarzschild metric rather than the McVittie metric, at large distances from the SMBH this coincides with the delay predicted at the beginning of this Section, Eq.~\eqref{eq:timing}.}, this is given by~\cite{Weinberg:1972kfs,Will:2014kxa}
\begin{align}\label{eq:shapiro}
    d_S(t) = r_s (1+ \bar z_s) \; &\left( \ln ( \frac{ \sqrt{b^2  + R^2} + R }{b}) \right. \nonumber \\
     & \left. +\kappa \ln ( \frac{\sqrt{b^2 + r(t)^2} + \kappa  r(t)}{b} )
    \right) \nonumber \\
    \simeq  -r_s (1+ \bar z_s) \; & \ln (1 -r(t)/a_{\bullet}) + {\rm const.}
    \, .
\end{align}
where $\kappa = \pm 1$ depending on whether the source is behind or in front of the SMBH, respectively. 
Here $r_s = 2 M_\bullet $ is the Schwarzschild radius of the SMBH, $R$ is the distance between the observer and the SMBH and  
\be
b = \sqrt{a_{\bullet}^2 - r(t)^2} = a_{\bullet} \sqrt{1 - \cos^2 \iota_\bullet \sin^2  (\Omega_{\bar z} t+ \phi_{\bullet})} \; .
\ee
is the impact parameter of the unperturbed trajectory from the source to the observer. See again Fig.~\ref{fig:lensing_setup_new} to visualize the geometry of the system. In the second line of Eq.~\eqref{eq:shapiro}, we used the fact that $R \gg b$ and that its constant contribution can be re-absorbed in the definition of the time of coalescence. 

Equation~\eqref{eq:shapiro} shows that the Shapiro delay introduces a direct dependence on the SMBH mass $M_{\bullet, \bar{z}_s}$, through $r_s$. This breaks the degeneracy introduced with the Doppler delay~\eqref{eq:DDT} between the outer orbit inclination and the central mass. A detectable Shapiro delay will then allow LISA to measure all three environmental parameters (the SMBH mass $M_{\bullet, \bar{z}_s}$, inclination $\iota_{\bullet}$ and distance from the SMBH $a_{\bullet, \bar{z}_s}$), as we show in the parameter estimation example in Sec.~\ref{pe_agn}.

The Shapiro delay in our system happens to be numerically similar to the light crossing time of the Earth orbit, so its modulation of the GW phase is comparable to the Doppler modulation induced by the LISA spacecraft motion,
\begin{align}\label{deltaphi_S}
    \delta \phi_S &\simeq 2 \pi f (d_{S,\rm max} - d_{S, \rm min}) \nonumber \\
    &\simeq 62 \, {\rm rad} \, \left(\frac{f}{10\mathrm{mHz}}\right) \left(\frac{M_{\bullet, \bar z}}{10^8 M_\odot}\right) \log \left(\frac{1-\cos \iota_\bullet }{1+\cos \iota_\bullet }\right) \, .
\end{align}
The Shapiro effect is therefore subdominant compared to the Doppler effect in GW190521-like binaries in LISA, but still detectable, see also Fig.~\ref{fig:plot_DopShaonly}. As we will see in Section~\ref{pe_agn}, the Shapiro effect is strong enough to break the degenaracies between the outer orbital parameters and improve our ability to constrain the SMBH mass. 

From Eq.~\eqref{deltaphi_S}, we also see that the Shapiro phase modulation has a stronger dependence on the inclination of the outer orbit $\iota_\bullet$ compared to the Doppler effect, with the effect peaking when the orbit and the line of sight are aligned~\cite{Meiron:2016ipr}. This can be seen comparing the two panels in Fig.~\ref{fig:plot_DopSha}. For high alignments, however, strong lensing is not negligible, as we argue below.  

For completeness, we also write the Shapiro redshift {by differentiating Eq.~\eqref{eq:shapiro} and keeping track of all redshift factors,} 
\begin{align}
z_S 
=&r_s (1+\bar{z}_s) \frac{\Omega_{\bar z}  \cos \iota_\bullet \cos ( \Omega_{\bar z} t +\phi_{\bullet})}{ \left( 1 - \cos \iota_\bullet \sin ( \Omega_{\bar z} t +\phi_{\bullet})\right)} \, .
\end{align}
The Shapiro redshift {oscillates with amplitude $\delta z_S = z_{S,\rm max} - z_{S,\rm min}$}, 
\be
\delta z_S \simeq 10^{-4} \frac{\cos \iota_\bullet}{\sqrt{1-\cos^2 \iota_\bullet}}  \, \left(\frac{a_{\bullet, \bar z}}{700M_{\bullet, \bar z}}\right)^{-3/2} \left(\frac{M_{\bullet, \bar z}}{10^8 M_\odot}\right)^{3/2}  ,
\ee
suppressed compared to the oscillation in the Doppler redshift, at least at the generic inclinations considered in this work.

\subsection{Other effects}
In this work, we focus on systems dominated by the Doppler and Shapiro effects: GW190521-like binaries on a circular orbit of hundreds of Schwarzschild radii around a $10^8$ - $10^9 M_{\odot}$ SMBH.
In general, hierarchical triple systems can be affected by a number of other dynamical effects, which might leave a detectable imprint on the GW signal for more extreme system parameters~\cite{DOrazio:2019fbq,Randall:2019sab,Yu:2020dlm,Chandramouli:2021kts,Torres-Orjuela:2018ejx}. Here we briefly summarize these effects. 
 
 \subsubsection{Lensing}\label{sec:lensing}
We treat the SMBH as a point-like lens~\cite{DOrazio:2019fbq}, with the binary moving on a sphere of radius $a_{\bullet}$ centered on the lens, see Fig.~\ref{fig:lensing_setup_new}. Projected along the line of sight, the angular-diameter distance between the observer and the SMBH and the observer and the source are denoted by $D_L$ and $D_S$, respectively. The angular-diameter distance between the lens and the source, $D_{LS}$, can take either sign during the evolution of the binary on the outer orbit, and coincides with $r(t)$ defined in the previous Section. In this lensing system, the Einstein radius is given by (assuming $D_{S}\sim D_L$ as $a_{\bullet} \ll D_L$, {and $D_{LS}>0$}), 
\begin{align}\label{RRE}
r_E=&(2 r_s  D_{LS})^{1/2} \nonumber \\
=&(2 r_s a_{\bullet} \cos \iota_\bullet \sin (\Omega_{\bar z} t+\phi_{\bullet}))^{1/2}\,,
\end{align}
and its maximum is at $\Omega t+\phi_{\bullet}=\pi/2$
\be
r^{\text{max}}_E=(2 r_s  a_{\bullet} \cos \iota_\bullet )^{1/2}\,.
\ee

One can identify three lensing regimes based upon the motion of the source with respect to the lens over its observable time~\cite{DOrazio:2019fbq}. The three relevant scales are the orbital period of the outer binary $T$, the time in band $T_{\text{obs}}$ and the time for the binary source to cross the Einstein radius of the lens, 
\be
T_{\text{lens}}=r_E^{\text{max}}/v \simeq 0.02 \; {\rm yr} \; (\cos \iota_\bullet )^{1/2} \left(\frac{a_{\bullet}}{700 M_{\bullet}}\right) \left(\frac{M_{\bullet}}{10^8 M_{\odot}}\right) \, .
\ee
The resulting three lensing regimes are 1) the repeating lens regime, for $T<T_{\text{obs}}$\,; 2) the slowly-moving lens regime, $T\geq T_{\text{obs}} 	\geq T_{\text{lens}}$\,; 3) the stationary lens regime, $T\geq T_{\text{obs}} \leq T_{\text{lens}}$\,.
GW190521-like binaries with outer orbital period given by Eq.~\eqref{eq:outer_period} fall in the repeating lens or in the slowly-moving lens regime, depending on $T_{\text{obs}}$ and $M_{\bullet \bar z_s}$.

A significant lensing event occurs when the source passes within one Einstein radius of the lens. Therefore, the probability of observing a lensing event after observing the binary for an entire orbit is the ratio of the binary inclination angle for which the source falls within one Einstein radius of the lens to the total possible range of inclinations. This probability depends on three parameters: $T_{\text{obs}}$, $a_{\bullet}$ and $M_{\bullet}$. In Ref.~\cite{Toubiana:2020drf}, we found that the probability of lensing occurring during a LISA observation { was above $10\%$ in a good portion of the lens parameter space, e.g., for $a_{\bullet} \lesssim 200 M_\bullet$ for $M_\bullet=10^8$. However, in the remainder of this work, we decided to focus for simplicity on the most likely scenario, in which lensing (including ``retro-lensing"~\cite{Yu:2021dqx}) is suppressed. This occurs for most outer orbital inclinations $\iota_{\bullet}$, for which the source is always outside the Einstein radius of the lens:}
%lensing (including ``retro-lensing"~\cite{Yu:2021dqx}) will be suppressed and therefore  negligible for most inclinations $\iota_{\bullet}$.} 
%In the remainder of this work, we assume an outer orbital inclination large enough that the source is always outside the Einstein radius of the lens, so that we can neglect the effect of lensing: 
\begin{equation}
    |y_{\text{min}}| =  \left|\frac{b}{r_{\rm E}}\right|_{\text{min}} = \frac{a_{\bullet \bar z_s} |\sin(\iota_\bullet)|}{\sqrt{2 r_s (1+ \bar z_s) a_{\bullet \bar z_s} \cos(\iota_\bullet)}}  \gg 1 \, .
\end{equation}
Choosing $\iota_\bullet = \pm 30 $ deg, for example, corresponds to $ |y_{\rm min}| \simeq 10 \left(M_{\bullet,\bar z_s}/10^8 M_\odot \right)^{-1/2} \left(a_{\bullet \bar z_s} / 700 M_{\bullet, \bar z}\right)^{1/2} \gtrsim 1$. 
\subsubsection{Relativistic and tidal effects}\label{sec:other}

Other third-body effects include: the outer orbit eccentricity, the geodetic or de Sitter precession of the inner orbit around the orbital angular momentum of the outer binary; Kozai-Lidov { and other resonant} oscillations of the inner binary inclination and eccentricity; the Lense-Thirring precession of the inner orbit around the SMBH spin, and aberration. For a summary of potentially relevant time delays, see e.g.~\cite{Ben-Salem:2022txj}. 

To estimate the significance of these effects, one can compare their period to the observation time. While the Doppler and lensing/Shapiro effects oscillate with the orbital period, Eq.~\eqref{eq:outer_period}, the de Sitter precession period is~\cite{Yu:2020dlm}
\begin{equation}
T_{dS} = 8 \times 10^2 \, {\rm yr} \, (1-e^2) \left( \frac{M_\bullet}{10^8 M_{\odot}}\right) \left( \frac{a_{\bullet}}{700 M_\bullet}\right)^{5/2},
\end{equation}
while the Lense-Thirring precession period reads~\cite{Yu:2020dlm}
\be
T_{LT} = 7 \times 10^4 \,{\rm yr} \, (1-e^2)^{3/2} \chi_\bullet  \left( \frac{M_\bullet}{10^8 M_{\odot}}\right) \left( \frac{a_{\bullet}}{700 M_\bullet}\right)^3,
\ee
where we introduced the outer orbit eccentricity $e$ and the SMBH spin $\chi_\bullet$.
The Kozai-Lidov oscillations depend also on the GW frequency~\cite{Yu:2020dlm}, with a typical timescale given by
\be
T_{KL} = 5 \times 10^5 \,{\rm yr} \, (1-e^2)^{3/2}  \left( \frac{M_\bullet}{10^8 M_{\odot}}\right)^2 \left( \frac{a_{\bullet}}{700 M_\bullet}\right)^3 
\left( \frac{f}{10 \rm mHz}\right).
\ee
The estimates above show that higher order effects can be safely neglected in our systems, as they occur on a much longer timescale {-- unless the eccentricity of the outer orbit is high, which is not expected for binaries migrating in AGN disks}. 
Aberration~\cite{Torres-Orjuela:2018ejx,Torres-Orjuela:2020cly} is also suppressed by the ratio between the outer orbital period and the inner one (for the phase shift) and by the ratio between the orbital velocity and the speed of light (for the amplitude correction). { Apsidal precession resonance~\cite{Liu:2020tcd}, which can drive eccentricity oscillations in the inner binary, are also negligible when the outer orbit is circular (as assumed in this work in view of the circularizing effect of the AGN environment).}

 \section{Parameter estimation in AGN}\label{pe_agn}
We now investigate the impact of a central SMBH on the GW emission from GW190521-like binaries detected by LISA, using the waveform model described in the previous Section.
We consider a GW190521-like binary located close to the SMBH (as suggested by Ref.~\cite{PhysRevLett.124.251102}), namely at a distance $a_{\bullet}=700 M_\bullet$
 for an observed SMBH mass $M_\bullet=10^8M_{\odot}$ {(see also Table~\ref{tab:pe_par_agn})}. In this Section, we drop the redshift subscript from all quantities, for simplicity. {We set the inital phase of the outer orbit to $\phi_{\bullet}=- 3\pi/4$, so as not to produce a fine-tuned enhancement or suppression of the environmental effects in the limit of large orbital separation.}
 In this exercise, we neglect the effect of the AGN disk gas and concentrate only on the Doppler modulation of the GW signal due to the orbital motion of the binary around the SMBH, as well as the time-dependent Shapiro delay. We choose a generic inclination of the orbit with respect to the line of sight, so that lensing is negligible. We also neglect other possible interactions of the binary with the SMBH that are suppressed at intermediate separations (Kozai-Lidov resonances,  orbital precession, etc.). 
 
{\centering
  \begin{table}[t] 
  \begin{ruledtabular}
   \begin{tabular}{cccc}
   $a_{\bullet}$ & $M_{\bullet} [M_{\odot}] $ & $\iota_{\bullet} $ [deg] & $\phi_{\bullet} $ \\
   \hline
   700 $M_{\bullet}$ & $10^8 $  & 30 & $-3\pi/4$ 
 \\
   \end{tabular}
   		\end{ruledtabular}
\caption{ The {(cosmologically redshifted)} parameters of the orbit around the AGN and the mass of the SMBH used in the parameter estimation example, inspired by the candidate electromagnetic counterpart of GW190521~\cite{PhysRevLett.124.251102}.
}\label{tab:pe_par_agn}
\end{table}}
 
 \renewcommand{\arraystretch}{1.5}
{\centering
  \begin{table}[t] 
    \begin{ruledtabular}
   \begin{tabular}{ccccc}
   $\mathcal{M} [M_{\odot}]$ & $q$ & $\chi_{\rm PN}$ & $\chi_-$ & $d_L$ [Gpc] \\
   \hline
   104.28 & 1.68 &  -0.23 &  -0.23 & 1.4
 \\
   \end{tabular}
      		\end{ruledtabular}
\caption{The {(cosmologically redshifted)} parameters of the GW190521-like binary selected for parameter estimation, consistent with the posterior distribution of GW190521 as analyzed by the LVC~\cite{Abbott:2020mjq} and with SNR$=9.5$ in LISA. The other signal parameters are $\iota=0.85$ rad, $\lambda=5.668$ rad, $\beta= -0.15$ rad, $\psi=1.2$ rad, $\phi_{\rm obs}$. The initial GW frequency is $f_0=0.0061524060$, corresponding to $t_c=7$ yr in the vacuum injection. {\  The corresponding cosmological redshift is $z_s\simeq0.27$~\cite{Ade:2015xua}.}
}\label{tab:pe_par}
\end{table}}

  \subsection{Waveform model and data analysis methods}
  
As the source-frame GW model, we take PhenomD \cite{Husa2016,KHH16}, which describes the inspiral-merger-ringdown signal for spin-aligned {circularized} binaries using only the dominant GW mode. For LISA observations, only the inspiral part of the signal is relevant, and precession and sub-dominant modes are severely suppressed by the post-Newtonian factor. The waveform is generated directly in the frequency domain.
 
 As described in details in Section~\ref{sec:theory}, the phase of the GW will be strongly modulated by the Doppler and Shapiro effects, so that the signal will appear as {chirping (when the binary is behind the SMBH) and anti-chirping (when the binary is in front of the SMBH).} The orbital period of our outer binary is about 1.8 years, and assuming 6 years of LISA observations and that the binary is observed 7 years before the merger, we expect to see approximately 3 full orbital cycles. One can clearly see the three orbital cycles and two (anti)chirping phases in the signal in Fig.~\ref{fig:plot_DopSha}. 
 
We transform the source-frame signal into the time domain, then model the phase modulation as a time-dependent delay, given by Eqs.~\eqref{eq:timing}, \eqref{eq:DDT} and \eqref{eq:shapiro}. We then split the signal into chirping and anti-chirping parts and transform the signal back into frequency domain using the SPA. Our piece-wise SPA is an improvement over the global approximation adopted, e.g., in Refs.~\cite{Yu:2020dlm,Yu:2021dqx}.
Our approximation only breaks at the turning points, which would require a higher order SPA~\cite{Hughes:2021exa}. We decided to neglect these parts of the signal as they contribute only a very small fraction of the total SNR. 

 In Section~\ref{sec:theory}, we also argued that the amplitude of the GW will receive a correction and oscillate with time as a result of the peculiar motion.
 This amplitude correction has a more modest effect on the waveform compared to the phase modulation: the overlap between a signal with and without amplitude correction is 0.9998 for our system parameters (and a signal with SNR$=9.3$). Neglecting the amplitude modulation should introduce a bias of approximately~$2 \%$ in the luminosity distance, much smaller than the measurement error for this parameter. For this reason, we neglect the amplitude correction and only model the time delay.

 Before we proceed to parameter estimation, we introduce the signal parametrization. For the vacuum waveform, we use the chirp mass $\mathcal{M}_c$, the mass ratio $q=m_1/m_2 \geq 1$, and the following combination of physicals spins $\chi_{PN} = \eta \left[ \left( 113 q+ 75 \right)\chi_1  + \left( 113/q + 75\right)\chi_2 \right]/113$,
 $\chi_{-} = q/(1+q) \chi_1 - 1/(1+q) \chi_2$,  
 where $\eta = q/(1+q)^2$ is the symmetric mass ratio and $\chi_{1,2}$ are spins of two BHs. The other vacuum parameters are standard: the luminosity distance $d_L$, the sky position as ecliptic longitude and latitude $(\lambda, \beta)$, inclination $\iota$, the polarization $\psi$ and the azimuthal position of the observer $\phi_{\rm{obs}}$. An important characteristic of the system is the time to merger, which we parametrize through the initial GW frequency $f_0$, which is the {observed} GW frequency (uniquely defined in the absence of higher order modes and precession) at the start of LISA observations, and kept the same when we place the binary around the SMBH or in vacuum. 
 The AGN-related parameters we use to parametrize the signal are the orbital (Keplerian) angular velocity $\Omega$, the projections of the orbital radius $a_\bullet \cos{\iota_\bullet}, \; a_\bullet \sin{\iota_\bullet}$ and initial phase $\phi_{\bullet}$. 
 
{For the GW190521-like binary that we use as a representative of the AGN binary population, we choose the parameters given in Table~\ref{tab:pe_par}. This particular choice of the parameters, over all the possible values given by the LVC posterior samples for GW190521~\cite{Abbott:2020mjq}, was made to maximize the vacuum SNR. For the same reason, we also set $t_c=7$ years. }

We perform parameter estimation using parallel tempering Markov-chain Monte-Carlo within a Bayesian framework, see~\cite{Marsat:2020rtl,Toubiana:2020cqv} for details. The signal we have chosen produces SNR$=9.5$ in vacuum, 
 (SNR$=9.3$ in the presence of the SMBH) in the LISA band using the noise budget outlined in the LISA science requirement document, SciRD~\cite{LISAscience_document}.
 
\subsection{Measurement of the central black hole orbit and properties}
\begin{figure}[t]
\centering
\includegraphics[trim={18.cm 0.cm 14.5cm 0cm},clip,width=.99\columnwidth]{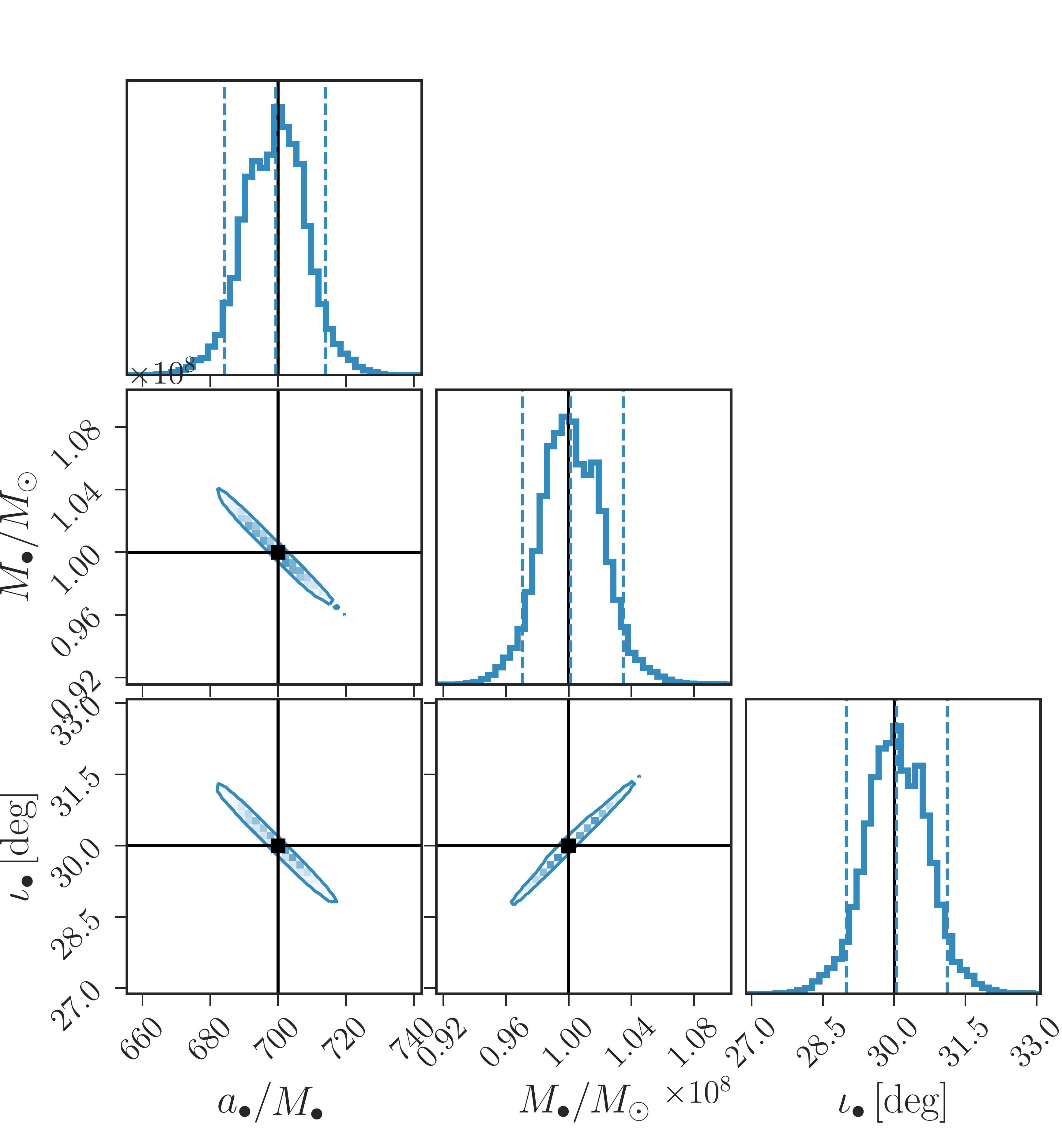}
\caption{Inference of the SMBH mass ($M_\bullet$) and the parameters of the outer orbit: inclination $\iota_{\bullet}$ and
orbital radius $a_{\bullet}$. The true {(cosmologically redshifted)} parameters are marked by black lines, while the dashed vertical lines indicate 90\% CLs.}  \label{F:AGN_params}
\end{figure}
The complete results of the parameter estimation are given in Appendix~\ref{app:fullPE}. Here we focus on the most interesting system parameters, marginalising over the others. 

The combination of a very strong Doppler modulation and a time-varying Shapiro delay allows us to constrain the orbital parameters of the outer binary and the mass of the AGN BH, as shown in Fig.~\ref{F:AGN_params}. For this particular system, we can determine the mass of SMBH to about 8\% level and the orbital inclination within a few degrees,  {together with the separation of the outer orbit within $3 \%$}. Comparably precise SMBH mass measurements are currently possible only for SgrA$^*$ \cite{2022A&A...657L..12G}, M87 \cite{2019ApJ...875L...6E}, and an handful of galaxies with detected nuclear megamaser emission \cite{2011ApJ...727...20K}. 
{GW190521-like binaries detected by LISA would therefore provide a competitive, complementary opportunity to measure SMBH masses through GW observations.}
{Electromagnetic- and GW-based measurements would be affected by different systematic uncertainties, which could help strengthen the measurement {accuracy}.}
 
\begin{figure*}[!htb]
\centering
\includegraphics[width=.95\textwidth]{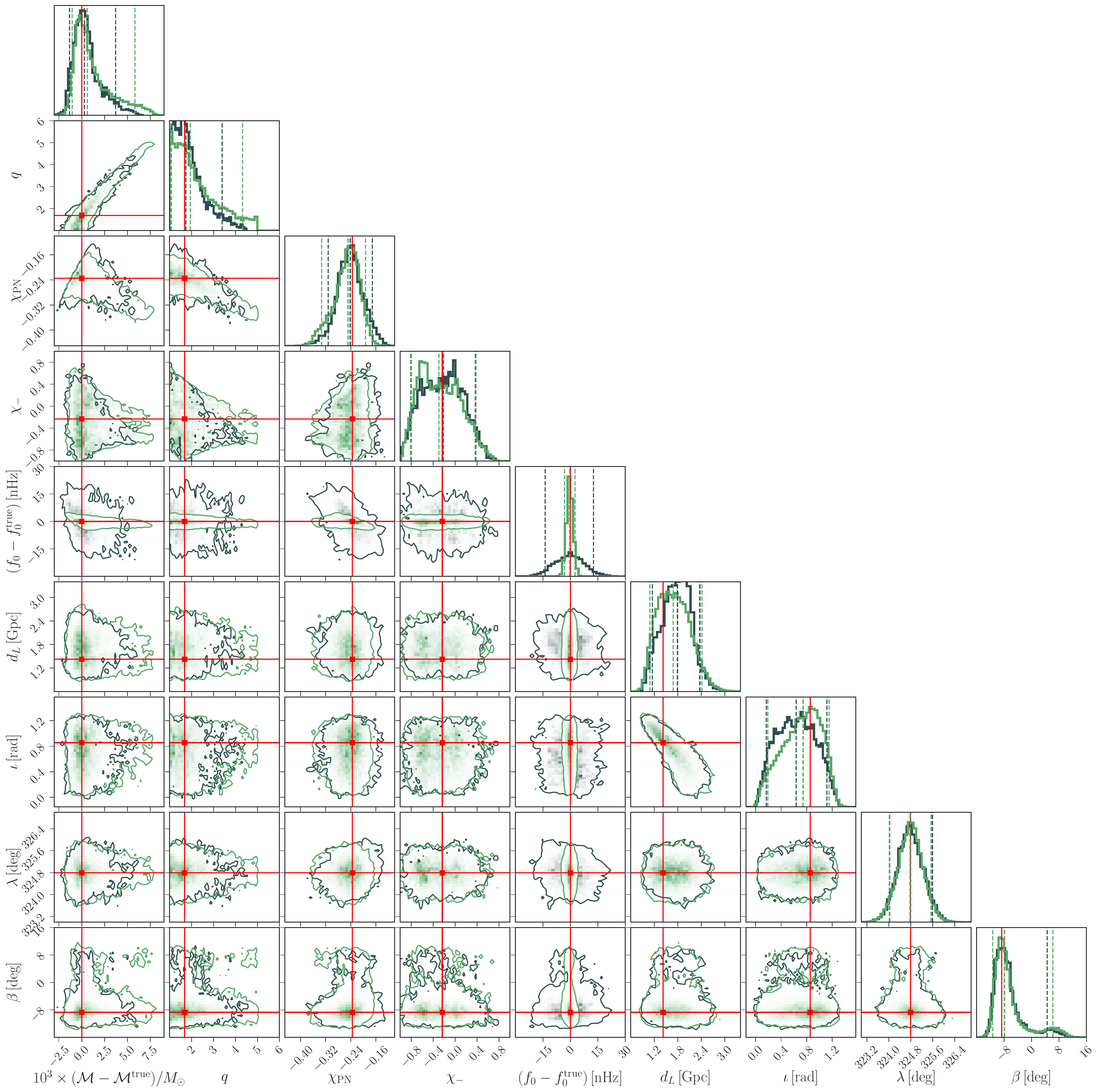}
\caption{Overplotting the uncertainties contours in the parameters of the GW190521-like binary (excluding AGN-related parameters) in the presence of the SMBH (green) and in its absence (in black). Red crosses denote the position of the true {(cosmologically redshifted)} parameters{, while the dashed vertical lines indicate 90\% CLs.}.}
\label{F:ZTF_gr_pe}
\end{figure*}

The Doppler modulation of the phase of the GW signal is a strong effect, but it is mostly orthogonal to the vacuum phase evolution; as a result, the parameters of the AGN orbit are uncorrelated to the parameters of the emitting binary. 
To show this explicitly we overplot in Fig.~\ref{F:ZTF_gr_pe}, for the vacuum parameters, the posteriors obtained in the presence and in the absence of the SMBH. The only parameter for which the posterior distribution is affected by the presence of the SMBH is the initial GW frequency: the difference between vacuum (green contours) and the binary orbiting the SMBH (grey contours) is clearly seen in the figure. Note that the initial frequency is the \emph{observed} one and, therefore, the initial \emph{orbital} frequencies of the inner binary in the two cases are quite different, due to the Doppler shift caused by the orbital motion around the SMBH. {The two binaries therefore merge at different times, although the shape of the posterior for the time of coalescence is not strongly modified by the presence of the SMBH.} 

 \subsection{Wider orbits and detectability}
 \begin{figure*}[!htb]
\centering
\includegraphics[width=.95\textwidth]{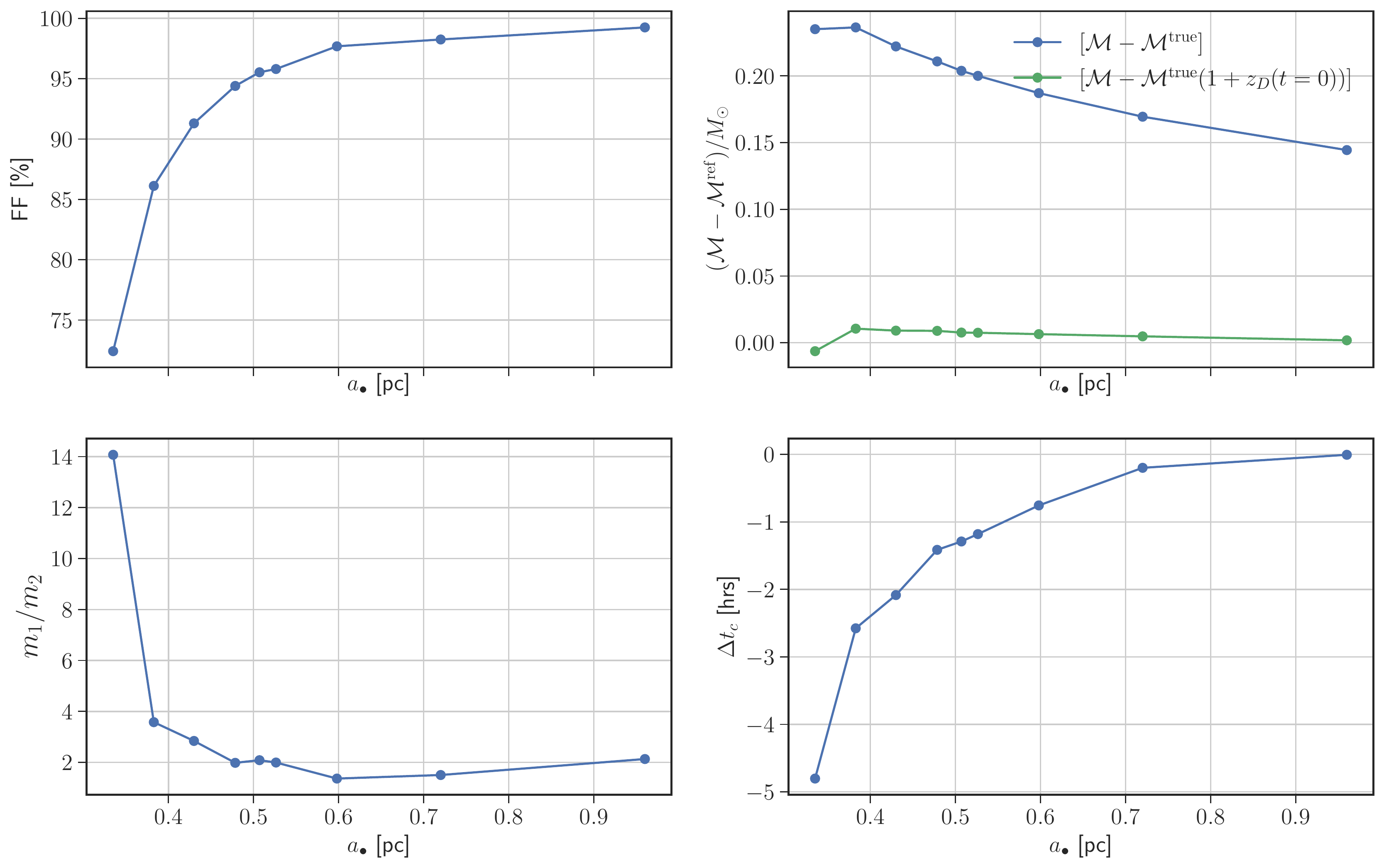}
\caption{Analysis of a GW190521-like binary at large distance from the SMBH with a vacuum template.  We show the fitting factor (upper left), the difference between best-fit and true {(cosmologically redshifted)} chirp mass (upper right), best-fit mass ratio (lower left) and difference between the best-fit and true {(observed)} time to coalescence (lower right), as a function of the distance of the GW190521-like binary from the SMBH. 
For the chirp mass, we also show the difference between the best-fit value and the true value redshifted by the Doppler effect, as measured at the start of observations.
As the distance from the SMBH decreases, the vacuum template causes a significant loss in SNR and strong biases in the inner binary parameters. The true value for the chirp mass and mass ratio is $104.28\,  M_{\odot}$ and $1.68$, respectively, the measurement precision (half of the 90\% confidence interval in Fig.~\ref{F:AGN-pe}) is approximately~$5 \times 10^{-3} M_{\odot}$ for the chirp mass and 20 minutes for the time to coalescence. 
}
\label{F:Detect}
\end{figure*}
{As we move the binary further away from the SMBH by increasing the size of the orbit, we approach the regime where the acceleration projected on the line of sight is constant, presented in our previous
publication \cite{Toubiana:2020drf}. In this limit, the acceleration enters the GW phase at -4PN, a term otherwise absent in vacuum templates. For even broader orbits, we reach the regime where {the acceleration is negligible and velocity is (almost) constant}. 
The latter is (almost) degenerate with a small modification in the redshift of the source, causing a slight shift in the observed (redshifted) masses of the binary. Therefore, for sufficiently wide orbits, we expect a vacuum template to be a good approximation of the signal, and a good template to detect the source with matched-filtering. 
}

To investigate the detectability of such a system using vacuum templates, we compute the fitting factor (i.e., the overlap maximized over all parameters). The fitting factor (FF), or rather $(1 - {\rm FF})$, gives the fractional loss in SNR due to mismatch between the best-fitting model and the signal \cite{Damour:2000zb}. The fitting factor takes into account that we still might match the signal at the expense of a bias in the parameter space, which can (at least partially) compensate for the mismatch in the model. 

{
We start with the binary orbiting the SMBH on a very broad orbit ($a_{\bullet}\simeq 1 $ pc), where the orbital speed is almost constant and can be re-absorbed into the redshift, with FF close to $100$\%. In Fig.~\ref{F:Detect}, the top left panel shows the evolution of the FF as we bring the binary closer to the SMBH. 
The top right plot gives the value of the chirp mass which maximizes the overlap between the signal and the model. In particular, we show the difference between the best-model chirp mass and either 1) the true (cosmologically redshifted) chirp mass or 2) the true chirp mass corrected by the Doppler redshift computed at the start time of observations.
The lower row of plots shows the best-model mass ratio (bottom left) and the difference between the best-model time of coalescence $t_c$ and true value in the observer frame (bottom right). }

{
The bias in the parameters recovered by our search has a clear increasing trend as we decrease the separation from the SMBH. At the closest distance we explored, $a_{\bullet} \lesssim 0.35$ pc, the recovered parameters (most noticeably, the mass ratio) display a jump. This is most likely due to non-optimal recovery of the FF maximum. Below this distance, the search for the FF maximum (in an 11-dimensional parameter space, although not all parameters are equally important) becomes challenging. 
}

{
We find that the bias in the chirp mass, although small in an absolute sense, is quite large compared to the typical measurement error for this parameter ($10^{-3} M_{\odot}$, see Fig.~\ref{F:ZTF_gr_pe}). At these large separations, however, most of the bias can be explained in terms of a constant Doppler redshift, as can be seen comparing the two curves in the top right panel of Fig.~\ref{F:Detect}.
}

{
The bias in the coalescence time is also large compared to the precision with which this parameter is measured -- $\mathcal{O}(20)$ minutes, worse than for typical stellar-origin BH binaries in LISA~\cite{Toubiana:2020cqv}. If such a binary was detected and analysed in LISA with vacuum templates, the bias in the coalescence time could prevent the association with the merger part of the signal detectable by ground-based detectors (unless other source parameters, such as chirp mass and sky position, already establish an association). {A bias in the merger time would also affect tests of GR, as emission into non-GR polarizations can also contribute to a shift in $t_c$ (see Ref.~\cite{Barausse:2016eii}) and thus be degenerate with the environmental effect studied here.}

We should stress that large biases might affect only a small fraction of AGN binaries, whose separation from the SMBH is sufficiently large for vacuum-template detection, but small enough to induce a sizeable bias. The severity of this problem for LISA will depend on how binaries are distributed within AGNs, and on the details of multiband strategies implemented by both space-borne and ground-based detectors in the future. 
}

We also expect the luminosity distance of the source to be significantly biased~\cite{Tamanini:2019usx} away from its background, cosmological value, with part of the bias explained by a constant Doppler redshift. However, since we neglected subleading amplitude modulations induced by peculiar motion in producing our signal, we cannot make quantitative statements on the luminosity distance bias. This is left for future work.

The trend in the FF shows that we start losing SNR for our GW190521-like binary at a distance $\approx 0.3-0.4$ pc: this is where the orbital acceleration becomes non-negligible, in agreement with our previous results~\cite{Toubiana:2020drf}. At these distances, a template including a -4PN term would be able to detect the source with higher SNR.

The trend in Fig.~\ref{F:Detect} indicates that we could not detect GW190521-like binaries around AGNs using pure vacuum GR templates for outer orbital radii below $0.3$ pc. Binaries on tighter orbits could only be recovered using the modified model described in this manuscript, including the full Doppler and Shapiro modulations. 
{ The use of these templates would increase the size of the template-bank to scan in matched-filtering by the number of independent outer orbit parameters (three), and thus increase the threshold required to claim a detection~\cite{Moore:2019pke}.
}
In practice, this means that we will most likely need to perform an ``archival search" \cite{Ewing:2020brd}, that is, to re-analyze LISA data if a massive binary, not identified in the online analysis, is detected by the 3rd generation of ground-based GW detectors (such as Einstein Telescope and Cosmic Explorer). A ground-based detection would narrow down the search in the parameter space and allow for the use of modified (non-vacuum GR) templates, like the one presented here.

\section{Conclusions}\label{sec:concl}
Interactions between BH binaries and the surrounding astrophysical environment are expected to have a negligible effect on the GW signals observable by LIGO and Virgo (although they may produce detectable electromagnetic counterparts~\cite{PhysRevLett.124.251102}). The reason is that these environmental effects tend to become important at lower frequencies, where they have time to build up a more significant phase difference in the gravitational waveforms.

From this point of view, it is quite natural to focus our attention on stellar-origin BH binaries in the lower frequency ($\sim$ mHz) LISA band. These systems are expected to be observable by LISA months or even years before they merge in the band of ground interferometers~\cite{Sesana:2016ljz}, and could probe their local environment, revealing the presence of gas in high densities or third bodies, possibly pointing at the formation mechanism of the binary itself or probing otherwise unexplorable galactic centers~\cite{Caputo:2020irr,Yu:2020dlm,Toubiana:2020drf}. The same systems could provide important information on putative low-frequency, non-GR effects, such as BH ``hair'' and vacuum dipole emission~\cite{Barausse:2016eii,Toubiana:2020vtf}. It is therefore crucial to understand how matter and the environment affect these binaries, both because interesting in themselves and because they may bias tests of GR.
 
 Recently, LIGO and Virgo have detected a particularly massive BH binary, GW190521, with total mass in excess of 100 $M_\odot$~\cite{Abbott:2020tfl,Abbott:2020mjq}. Such a large mass, as well as the possibility that this system may have an optical counterpart~\cite{PhysRevLett.124.251102}, may suggest that this binary formed in a gas-rich environment, e.g. an AGN disk, and that it might the first of many AGN (or GW190521-like) GW events.

 In this work, we have extended our earlier analysis of GW190521-like binaries in LISA~\cite{Toubiana:2020drf} to include two effects: the Doppler and Shapiro effects induced by the galactocentric motion of the binary. We proposed a way of producing accurate frequency-domain waveforms, combining the chirping and anti-chirping parts of the Doppler-modulated signal. By performing a full Bayesian analysis, we have found that 
 the Doppler modulation and Shapiro time delay can induce significant effects on LISA waveforms if the binary is at  distances of about a 
 thousand gravitational radii (or less) from the AGN's supermassive BH. We also show that these effects permit  breaking the degeneracy between purely vacuum waveforms and ones including environmental effects, thus allowing LISA to measure the AGN parameters with high precision (galactocentric radius of the binary, central BH mass and AGN inclination angle). Not only does this have important implications for AGN astrophysics, but it will also allow for potentially disentangling environmental imprints on the waveforms from putative non-GR effects entering in the same low-frequency band.
 
 {
In our analysis, we focused on the accurate treatment of the leading effects of the third body on the waveform, by focusing on the induced time shifts. While subdominant, amplitude effects, neglected here, should eventually be included in the waveform model to avoid minor biases in luminosity distance measurements (see e.g. \cite{Tamanini:2019usx} and \cite{Cusin:2020ezb}) and to improve the determination of the environmental parameters.  
 
 Another natural extension of our work will be to consider galactocentric orbits highly aligned with the line of sight, for which the effect of gravitational lensing is non-negligible. Our accurate treatment of Doppler modulated signals will allow for a better assessment of the detectability of lensing in AGNs, which in turn might improve the precision with which LISA could measure the SMBH mass. 
 Our method to treat the Doppler-induced signal modulation could also be used for systems on even tighter orbits around SMBHs, to better assess higher order effects relevant in this regime~\cite{Yu:2020dlm,Yu:2021dqx,Torres-Orjuela:2020cly}. 
 }
 
\acknowledgements

{We thank C.~Bonvin for helpful discussions.}
E.~B.~acknowledges financial support provided under the European Union's H2020 ERC Consolidator Grant ``GRavity from Astrophysical to Microscopic Scales'' grant agreement no. GRAMS-815673. This work was supported by the EU Horizon 2020 Research and Innovation Programme under the Marie Sklodowska-Curie Grant Agreement No. 101007855.
The work of G.~C.~is supported by Swiss National Science Foundation (Ambizione Grant).
S.~B., S.~M. and N.~T.~acknowledge support form the French space agency CNES in the framework of LISA. A.~S.~acknowledges financial support provided under the European Union’s H2020 ERC Consolidator Grant ``Binary Massive Black Hole Astrophysics'' (B Massive, Grant Agreement: 818691). A.~C.~is supported by the Foreign Postdoctoral Fellowship Program of the Israel Academy of Sciences and Humanities and also acknowledges support from the Israel Science Foundation (Grant 1302/19), the US-Israeli BSF (Grant 2018236), the German-Israeli GIF (Grant I-2524-303.7) and the European Research Council (ERC) under the EU Horizon 2020 Programme (ERC-CoG-2015-Proposal n. 682676 LDMThExp).
This project has received financial support from the CNRS through the MITI interdisciplinary programs.

\appendix
\section{Doppler effect in the large separation limit} \label{app:peculiar}

In this Appendix, we {show how the Doppler effect reduces to a {constant} peculiar velocity and peculiar acceleration (discussed in detail in Ref.~\cite{Tamanini:2019usx} ) in the large separation limit.} For simplicity, we neglect the cosmological background here. 
When the outer orbit is very wide, the standard SPA can be used to approximately compute the waveform in the frequency domain. Formally, this corresponds to expanding up to leading order in the separation of timescales between radiation reaction and outer orbital motion.
We can then write~\cite{Marsat:2018oam,Marsat:2020rtl},
\begin{align}
\label{eq:leadingtransfer}
    \tilde{h}_{\rm o}(f) \simeq &\int dt \; h(t+d(t)) \exp \left[ 2 \pi i  f t \right]  \nonumber \\
    \simeq &\tilde{h}(f) \exp \left[ -2 \pi i  f d_D(t_f) \right] \nonumber \\ 
    = &A(f) \exp \left[-i \varphi(f)\right] \exp \left[ -2 \pi i  f d_D(t_f) \right]   \,,
\end{align}
where we separated the vacuum signal $\tilde{h}(f)$ from the contribution due to the change in the time of arrival of the signal.
At leading order the time-frequency correspondence is given by the SPA for the vacuum signal only\footnote{The time is often defined with respect to the (constant) time of coalescence: $t_c-t_f = \frac{1}{2\pi} \frac{d \varphi}{d f}\sim - t_f $. We choose here a different convention, with the time measured from the start of the observation, $t_0=0$. This is more appropriate for binaries that do not merge within the observation time or in band.},
\be\label{eq:deftf}
    t_f = \frac{1}{2\pi} \frac{d \varphi}{d f} \,,
\ee
with $\varphi$ the Fourier domain phase of the vacuum signal. At leading post Newtonian (PN) order,
\be\label{eq:PsiV}
    \varphi(f) =  - \frac{3}{128 \eta} v^{-5} \,,
\ee
where $\eta$ is the symmetric mass ratio of the inner binary, we ignored a constant phase and a time shift, and $v \equiv (\pi M f)^{1/3}$. 
Notice that the frequency here is always the observed one, not that at the source.

Inserting Eq.\eqref{eq:PsiV} in Eq.\eqref{eq:deftf} one gets
\be
    t_f = \frac{5}{256 \eta} M v^{-8} \,.
\ee
At larger separations, we can expand the projected motion \eqref{eq:rt} in terms of the velocity and acceleration along the line of sight (pointing from the observer to the source) at $t=0$,
\be
    d_D(t) = r(t) =  \vv^\shortparallel_0  t + \frac{1}{2} a^\shortparallel_0 t^2 + \mathcal{O} (t^3) \, ,
\ee
and ignore a constant $r_0$, which defines the distance to the source in the vacuum waveform $\tilde{h}$. 
Then the observer-frame waveform can be written as $\tilde{h}_{\rm o}(f) = A_{\rm o}(f) e^{-i\varphi_{\rm o}(f)}$, with
\begin{align}
    \varphi_{o}(f) 
    &= - \frac{3}{128 \eta} v^{-5} \left(1 - \frac{5}{3} \frac{\vv^\shortparallel_0}{c} \right) + \frac{25}{65536 \eta^2} \frac{a^\shortparallel_0 M_z}{c} v^{-13} \,, \nn\\
\end{align}
where again we ignore terms that are constant and linear in the frequency, corresponding to a constant phase and a time shift. 
The first term corresponds to a Doppler shift and for a source at cosmological distance defines the redshift together with the usual cosmological contribution 
\be
    (1 + z) = (1 + \bar z_s) \left( 1 + \vv_0^\shortparallel/c \right) \,,
\ee
see also Ref.~\cite{Tamanini:2019usx}. 
The second term is a -4PN correction due to the peculiar acceleration, and was explored in detail in Refs.~\cite{Bonvin:2016qxr,Inayoshi:2017hgw,Tamanini:2019usx}. 

\section{Full posterior}\label{app:fullPE}
\begin{figure*}[!htb]
\centering
\includegraphics[width=\textwidth]{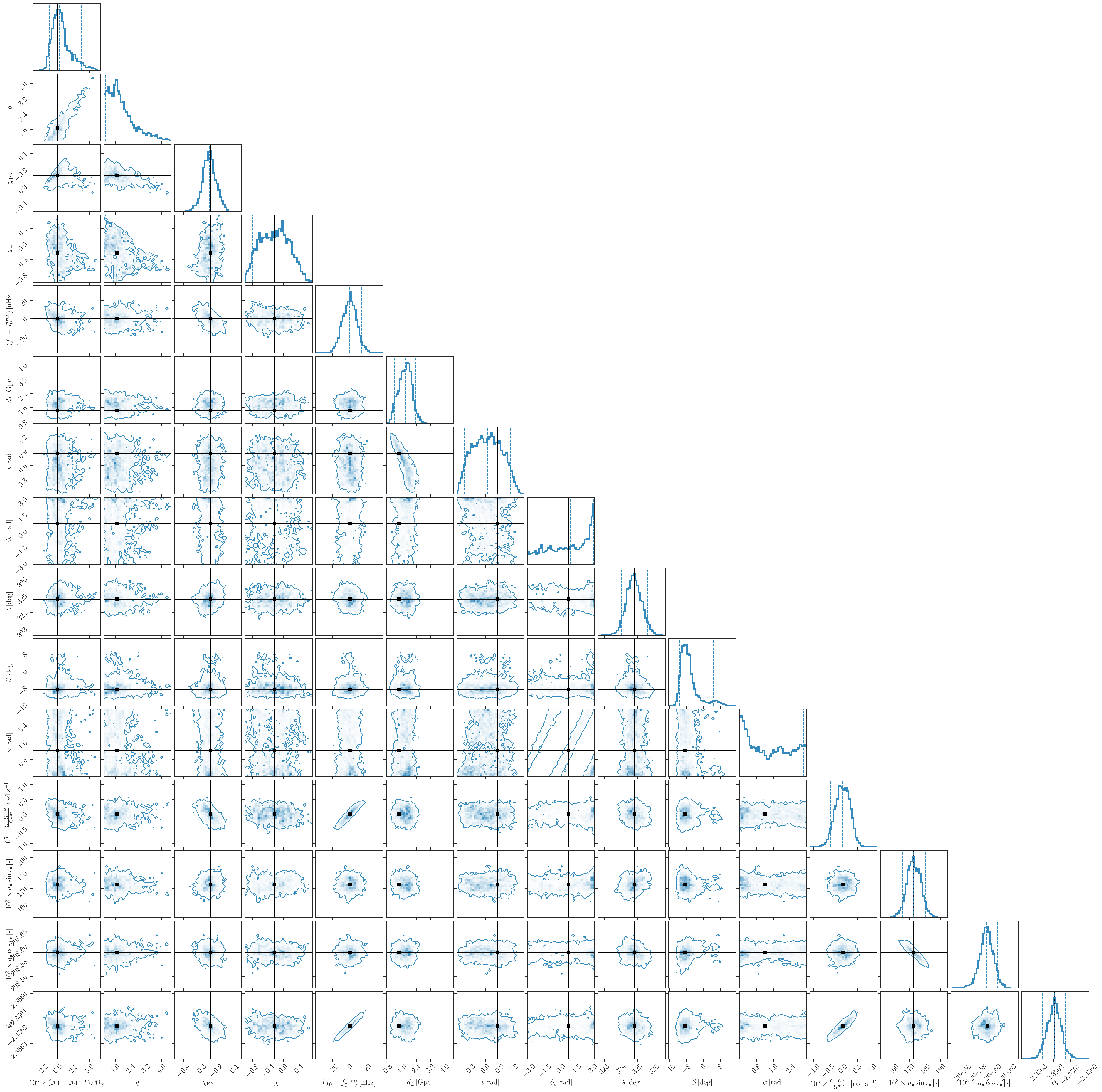}
\caption{Full parameter estimation of a GW190521-like binary around an AGN. The parameters are, from left to right: chirp mass, mass ratio, symmetric spin combination ($\chi_{PN}$), asymmetric spin combination  ($\chi_{-}$), initial GW frequency, luminosity distance,  azimuthal position of the observer in the source frame, ecliptic longitude and latitude, polarization phase, orbital angular velocity of the binary around AGN, projections of the orbital separation $a_\bullet \cos{\iota_\bullet}, \; a_\bullet \sin{\iota_\bullet}$, and initial centrogalactic orbital phase. The black lines mark the true parameters of the injected signal. } \label{F:AGN-pe}
\end{figure*}

The complete results of the parameter estimation of our prototype system are given in Fig.~\ref{F:AGN-pe}. The parameters related to the SMBH are given at the rightmost part of the corner plot. 
{
These results were confirmed with two independent samplers (our implementation of \texttt{emcee}~\cite{Foreman2013}, and one combining slice sampling with Metropolis-Hastings) and by performing several parameter estimation runs with different sampling settings. The results we present throughout this work were obtained with the best-performing sampler (\texttt{emcee}-like) and settings.
}

\bibliography{Ref}

\end{document}